\documentclass[usenatbib, fleqn]{aa}  
\usepackage{graphicx}
\usepackage{amsmath}
\usepackage{multicol}
\usepackage{color}
\usepackage{scrextend}
\usepackage[pdftex, linktocpage=true]{hyperref}
\hypersetup{
	colorlinks= true, 
	urlcolor= blue, 
	linkcolor=red, 
	citecolor=blue 
}
\usepackage{fancyhdr}
\usepackage{caption}
\usepackage{subcaption}
\usepackage{float}
\usepackage{wrapfig} 
\usepackage{pdflscape}
\usepackage{makecell}
\usepackage{amssymb}
\usepackage{booktabs}

\usepackage[flushleft]{threeparttable}
\newcommand\Tstrut{\rule{0pt}{2.6ex}}         
\newcommand\Bstrut{\rule[-0.9ex]{0pt}{0pt}} 
\usepackage{txfonts}
\usepackage{multirow}

\begin{document} 
	
	\title{Photoionisation modelling of the X-ray emission line regions within the Seyfert 2 AGN NGC 1068}
	
	\author{S. Grafton-Waters \inst{\ref{MSSL}} \and
		G. Branduardi-Raymont \inst{\ref{MSSL}} \and
		M. Mehdipour \inst{\ref{STSI},\ref{SRON}} \and
		M. Page \inst{\ref{MSSL}} \and
		S. Bianchi \inst{\ref{Roma}} \and
		E. Behar \inst{\ref{Israel}} \and
		M. Symeonidis \inst{\ref{MSSL}}
	}
	
	\institute{
		Mullard Space Science Laboratory, University College London, Holmbury St. Mary, Dorking, Surrey, RH5 6NT, UK \\ \label{MSSL} \email{sam.waters.17@ucl.ac.uk} \and 
		Space Telescope Science Institute, 3700 San Martin Drive, Baltimore, MD 21218, USA \label{STSI} \and
		SRON Netherlands Institute for Space Research, Sorbonnelaan 2, 3584 CA Utrecht, The Netherlands \label{SRON} \and
		Dipartimento di Matematica e Fisica, Università degli Studi Roma Tre, Via della Vasca Navale 84, 00146 Roma, Italy \label{Roma} \and
		Department of Physics, Technion-Israel Institute of Technology, 32000 Haifa, Israel \label{Israel}
	}

	\date{Received date / Accepted date}
	
	
	\abstract
	{}
	{We investigate the photoionised X-ray emission line regions (ELRs) within the Seyfert 2 galaxy NGC 1068 to determine if there are any characteristic changes between observations taken 14 years apart.}
	{We compared XMM-Newton observations collected in 2000 and 2014, simultaneously fitting the reflection grating spectrometer (RGS) and EPIC-pn spectra of each epoch, for the first time, with the photoionisation model, \texttt{PION}, in \texttt{SPEX}.}
	{We find that four \texttt{PION} components are required to fit the majority of the emission lines in the spectra of NGC 1068, with $\log \xi = 1 - 4$, $\log N_H > 26$ m\textsuperscript{-2}, and $v_{out} = -100$ to $-600$ km s\textsuperscript{-1} for both epochs. 
		Comparing the ionisation state of the components shows almost no difference between the two epochs, while there is an increase in the total equivalent column density. To estimate the locations of these plasma regions from the central black hole, we compare distance methods, excluding the variability arguments as there is no spectral change between observations. 
		Although the methods are unable to constrain the distances for each plasma component, the locations are consistent with the narrow line region, with the possibility of the higher ionised component being part of the broad line region; we cannot conclude this for certain, but the photoionisation modelling does suggest this is possible. 
		In addition, we find evidence for emission from collisionally ionised plasma, while previous analysis had suggested that collisional plasma emission was unlikely. However, although \texttt{PION} is unable to account for the \ion{Fe}{XVII} emission lines at 15 and 17 \AA, we do not rule out that photoexcitation is a valid processes to produce these lines as well.	}
	{NGC 1068 has not changed, both in terms of the observed spectra or from our modelling, within the 14 year time period between observations. This suggests that the ELRs are fairly static relative to the 14 year time frame between observations, or there is no dramatic change in the spectral energy distribution, resulting from a lack of black hole variability.}
	
	
	\titlerunning{Photoionisation Modelling of the Emission Line Regions in NGC 1068}
	\authorrunning{Grafton-Waters et al.}
	\maketitle
	
	\section{Introduction}
	\label{Sec:Intro}
	
	The unification model for active galactic nuclei \citep[AGN;][]{Miller1983, Antonucci1985} is able to explain the paradigm that some spectroscopic observations of galaxies display broad emission lines, while others do not. The reason, quite simply, has to do with the orientation and our line-of-sight (LOS) towards the AGN of interest. How we perceive to view the source is the cause of this observational difference, whereby type 1 AGN are viewed from `face on', allowing for both narrow and broad line emission to be seen, whereas type 2 AGN are seen from `side on', so only narrow emission is detected. In type 2 AGN, there is some obscuring material (known as the dusty torus) blocking our LOS to the nucleus, meaning emission from the broad line region (BLR) is invisible; only emission from the narrow line region (NLR) is seen. However, broad Balmer emission lines in polarised light were observed within NGC 1068. They were explained as being due to photons from the BLR scattered off free electrons or possibly dust, and, therefore, polarised, into our LOS \citep{Antonucci1985}.
	
	
	In addition, the covering factor of the dusty torus must also be considered in the unification model, as determining between type 1 or type 2 AGN also depends on the probability that a photon can escape the obscuring material \citep{Ricci2017}. This suggests that the larger the covering factor, the more clumps are present within the torus, resulting in a large optical depth such that the photons are less likely to escape. Therefore, the probability of a photon escaping depends on the size of the torus (given by the covering factor), increasing the likelihood of the AGN being obscured \citep{Ricci2017}.
	
	A further consequence of an obscuring torus means the soft X-ray spectra of type 2 AGN do not show absorption features, but they rather display strong emission features above an absorbed continuum. This can be explained by the ionisation cone model \citep[e.g.][]{Kinkhabwala2002}, suggesting that the outflowing wind, which absorbs so much of the X-ray flux seen in Seyfert 1 AGN, re-emits the X-rays through all angles relative to the direction of outflow, producing the emission features that are seen in the spectra of type 2 AGN. In other words, the emission seen in Seyfert 2 AGN is the re-emission of the plasma from the warm absorber (WA) wind in Seyfert 1 AGN \citep[e.g.][]{Kaastra2002, Kaastra2012, Behar2017, Mehdipour2018}, from the evidence that the ionic column densities of the emission lines were consistent with the values of the WAs in Seyfert 1 AGN \citep{Kinkhabwala2002}. 
	
	NGC 1068 is a heavily studied type 2 Seyfert galaxy, which supports perfectly the unification \citep{Antonucci1985} and the ionised cone \citep{Kinkhabwala2002} models, as only narrow emission lines are observed in its soft X-ray spectra. NGC 1068 is at a redshift of $z = 0.0038$ \citep{Huchra1999}, with a super massive black hole (SMBH) mass of $M_{BH} = 1.6 \times 10^7$ M\textsubscript{$\odot$} \citep[e.g.][]{Panessa2006}.
	
	The very first high resolution X-ray spectra of NGC 1068 were taken by XMM-Newton \citep{Kinkhabwala2002} and Chandra \citep{Brinkman2002, Ogle2003}, in the early 2000s, both showing many narrow emission lines originating from photoionised equilibrium (PIE) plasma. \cite{Kinkhabwala2002} found that excess emission in the resonance lines was explained by photoexcitation, rather than emission from collisionally ionised plasma. They suggested that the addition of plasma in collisionally ionised equilibrium (CIE) could explain the emission excess in the lower order resonance lines, but would enhance the other lines in the spectrum, in particular over predicting the intercombination and forbidden lines of the He-like triplets. \cite{Kraemer2015} reanalysed the 2000 RGS spectrum with a two component model, and found the emission lines were blueshifted by $\sim 160$ km s\textsuperscript{-1}, consistent with the optical [\ion{O}{III}] $\lambda$5007 results \citep{Crenshaw2010}. Both \cite{Brinkman2002} and \cite{Kraemer2015} suggested that the optical and X-ray emission line regions are part of the same outflowing wind, because they have consistent velocities; this is a general property of type 2 Seyfert AGN \citep{Bianchi2006}.
		
	
	Due to the spatial resolution of Chandra, \cite{Brinkman2002} and \cite{Ogle2003} were able to differentiate between two emission regions within the nucleus of NGC 1068, made up of a central, primary region and an off-centre, secondary region. The secondary region is a bright X-ray source that coincides with bright radio and optical emission, indicating a correlation between the X-ray and optical regions, which may arise from old jet material \citep{Wilson1987,Young2001}. The secondary region was found to have an ionic column density three times smaller compared to the primary region, and plasma diagnostics were unable to conclude if the emission was produced in PIE or CIE plasma \citep{Brinkman2002}.
	
	The hard (2 - 10 keV, and above) X-ray spectrum for NGC 1068 is also very complex \citep[see e.g.][]{Guainazzi1999,Bianchi2001} whereby the emission features can be explained by multiple components \citep[e.g.][]{Bianchi2001, Matt2002, Matt2004, Pounds2006, Bauer2015}. From NuSTAR observations, three reflection components were necessary to model the \ion{Fe}{K$\alpha$} line and Compton hump with column densities of $10^{29}$, $1.5 \times 10^{27}$, and $5 \times 10^{28}$ m\textsuperscript{-2}. The third component bridged the gap between the cold and warm reflectors which dominated the \ion{Fe}{K$\alpha$} emission line, and the other two components produced the Compton hump \citep{Bauer2015}.
	
	The Atacama Large Millimeter Array (ALMA) observations of NGC 1068 have enabled mapping molecular tracers to study both the obscuring torus, circumnuclear disk (CND) and star burst ring \citep[SBR; e.g.][]{Garcia-Burillo2014, Garcia-Burillo2016, Garcia-Burillo2017, Viti2014, Imanishi2018}. The outflow rate within the CND was found to be an order of magnitude larger than the star formation rate, suggesting the ionised gas outflow is AGN driven \citep{Garcia-Burillo2014}. The SBR is colder and less dense than the CND \citep{Viti2014} as it is much further out from the black hole \citep[$r_{CND} < 200$ pc, $r_{\textit{SBR}} \sim 1.3$ kpc;][]{Garcia-Burillo2017}. From ALMA observations, it was found that the torus, with gas mass $M_{gas}^{torus} \sim 1 \times 10^5 M_{\sun}$ and a distance of $R_{torus} \sim 3.5$ pc from the black hole, is inhomogeneous, with a strong non-circular molecular motion, possibly inclined at larger radii \citep{Garcia-Burillo2016}. 
	
	With infrared observations, hydrogen emission structures have been observed to infall towards the central SMBH, where one molecular cloud is close enough ($\sim 1$ pc) such that a tidally distributed, optically thick stream may possibly makeup the outer part of a clumpy, dusty torus \citep{Muller-Sanchez2009}. Evidence for this clumpy structure, in the high energy X-rays, was found during an unveiling event in August 2014 when a flux excess was observed above 20 keV, compared to December 2012 and February 2015 \citep{Marinucci2016}. This was caused by a decrease in column density by about $\Delta N_H \sim 2.5 \times 10^{24}$ cm\textsuperscript{-2}, which allowed \cite{Marinucci2016} to infer an intrinsic 2 - 10 keV luminosity of $L = 7 \times 10^{43}$ erg s\textsuperscript{-1} for the central, obscured AGN. 
	
	The goal of this work is to attain a self-consistent best fit photoionisation model for the 2000 and 2014 XMM-Newton observations of NGC 1068 in order to achieve an in depth understanding of the emitting plasma regions within this AGN. We wish to determine if there are any changes regarding the emitting plasma between the epochs, to then obtain other properties  which can be 	derived from our photoionisation modelling. In this paper, we analyse both the reflection grating spectrometer \citep[RGS;][]{denHerder2001} and European photon imaging camera pn-CCD \citep[EPIC-PN;][]{Struder2001} spectra simultaneously from each observation in great detail, with a sophisticated and self consistent photoionisation model. 
	
	The layout of this paper is as follows. In Sect. \ref{Sec:Data_Reduction} we describe how we reduced the data, while in Sect. \ref{Sec:SED} we explain how we set up the ionising SED, and how this differs from the observed continuum. Section. \ref{Sec:Data_Analysis} outlines how we construct the model to analyse the spectra, before presenting our spectral fitting results in Sect. \ref{Sec:Results}. We compare our findings from both epochs in Sect. \ref{Sec:Comp_Epochs}, and then examine the locations and thermal properties of the emission line regions in Sect. \ref{Sec:Discussion}. Finally, we present our conclusions in Sect. \ref{Sec:Conclusions}.
	
	\begin{table}
		\centering
		\caption{Observation log for NGC 1068 showing the observation ID, start date, duration, and the instruments used in this investigation. These observations are modelled in this paper, after combining each epoch together (see Sect. \ref{Sec:Data_Reduction} for details). 
		}
		\label{Table:Obs_Log}
		\begin{tabular}{c | c c c c}
			\hline
			\hline
			Obs  & Obs & Start & Duration  & Instrument \\
			Date& ID & Date & (ks) & Used \\
			\hline
			\multirow{2}{*}{2000}&	0111200101 &2000/07/29 & 42 & PN, RGS \\
			& 0111200201 & 2000/07/30 &46 & PN, RGS \\
			\hline
			\multirow{4}{*}{2014} & 0740060201 & 2014/07/10 & 64 & PN, RGS \\
			& 0740060301 & 2014/07/18 & 58 & PN, RGS \\
			& 0740060401 & 2014/08/19 & 54 & PN, RGS\\
			& 0740060501 & 2015/02/03 & 55 & RGS\\
			\hline
		\end{tabular}
	\end{table}
	
	\begin{figure}
		\centering
			\begin{subfigure}{0.95\linewidth}
			\includegraphics[width=1\linewidth]{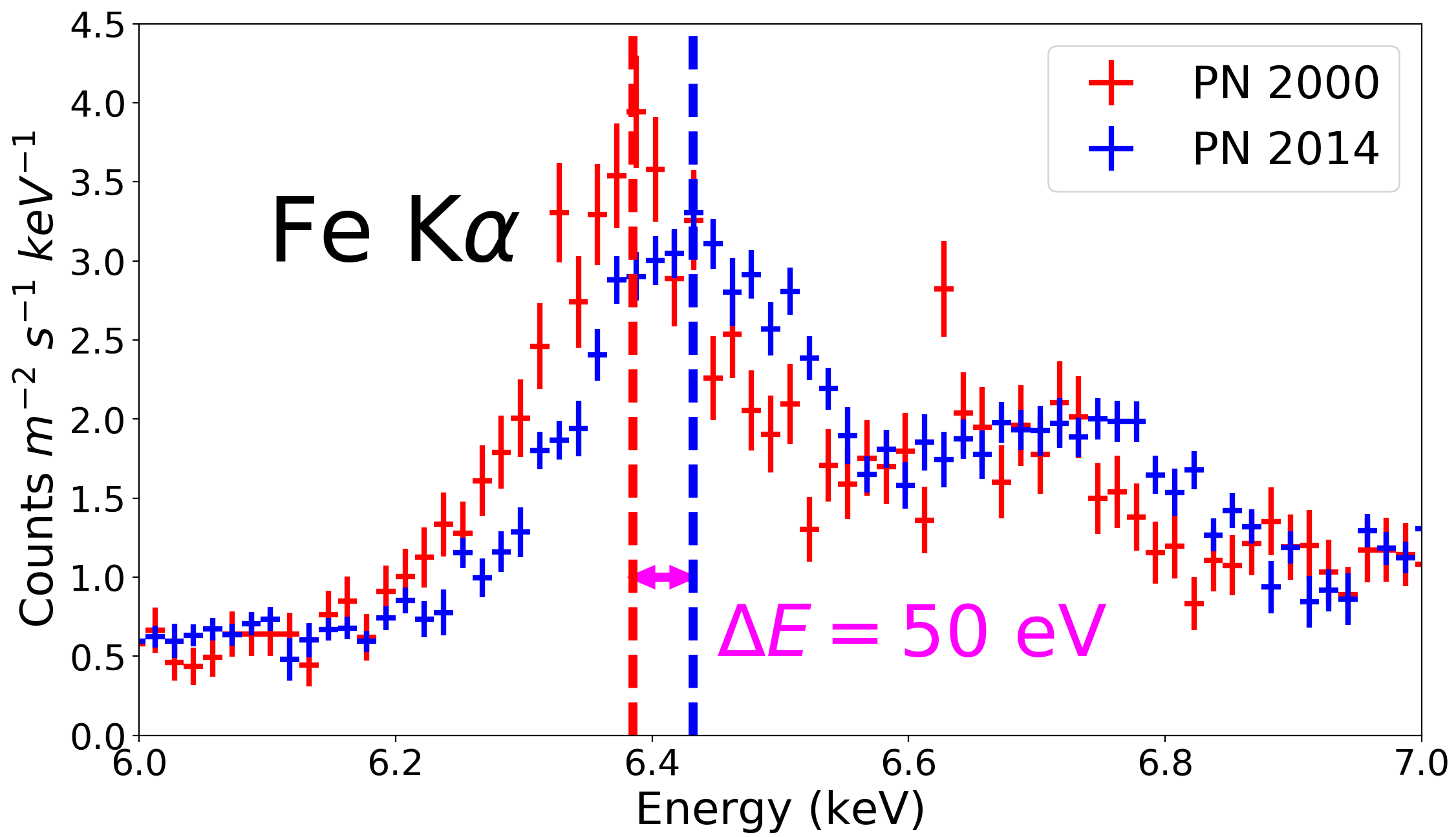}
		\end{subfigure}
		\begin{subfigure}{0.95\linewidth}
		\includegraphics[width=1\linewidth]{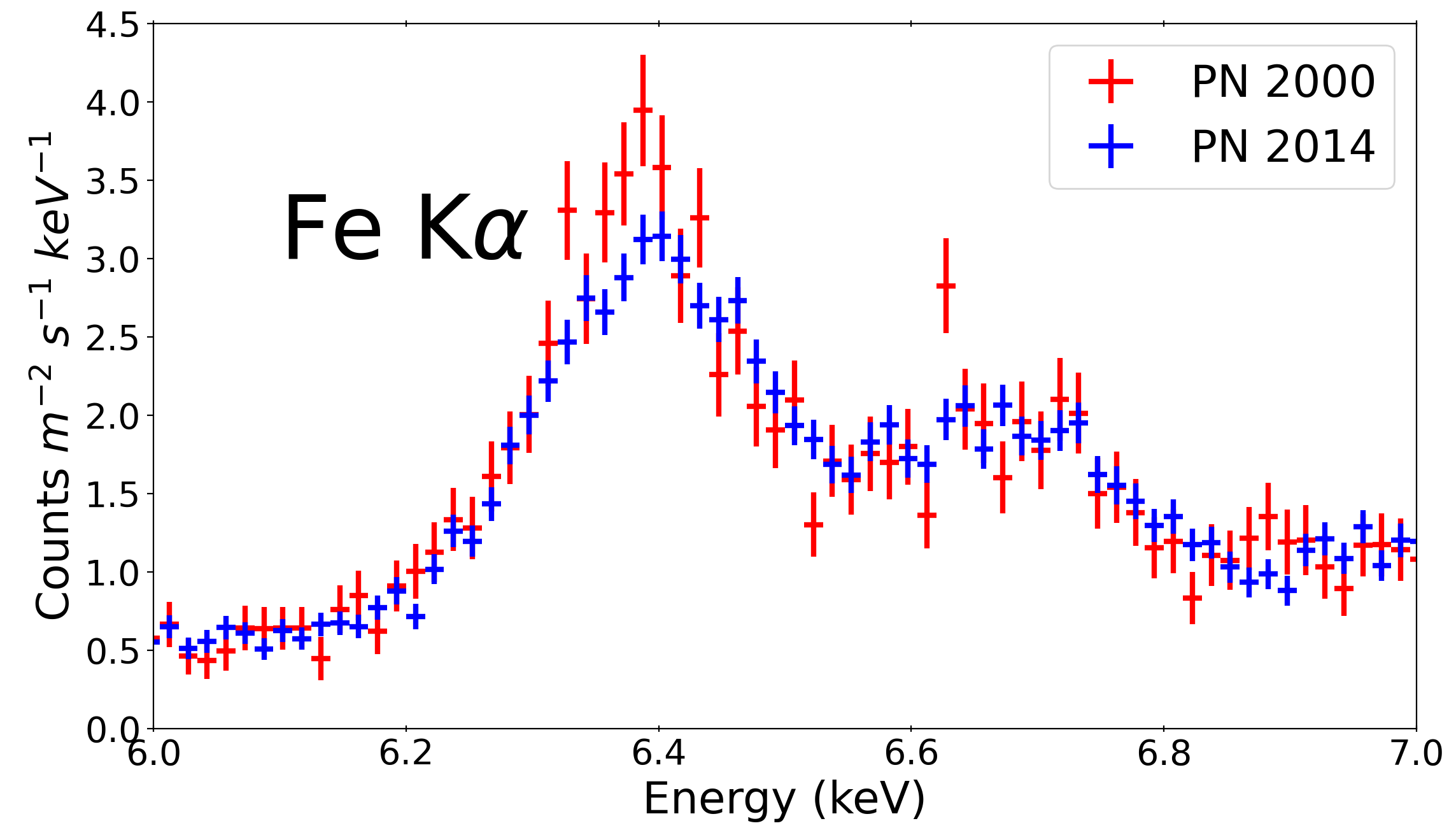}
		\end{subfigure}
		\caption{Comparing the \ion{Fe}{K$\alpha$} line in 2000 (red crosses) and 2014 (blue crosses). \textit{Top:} There is a clear shift in the \ion{Fe}{k$\alpha$} line between these two epochs, denoted by the red and blue dashed vertical lines at the peak energies of the observed data. The magenta arrow signifies this energy shift between the observations, equivalent to $\Delta E = 50$ eV. This shift is due to uncertainties in the long-term charge transfer inefficiency and instrumental gain \citep{Cappi2016}. \textit{Bottom:} We account for this shift manually by multiplying the energies (\texttt{PI}) of the event lists in each 2014 observation by 0.993. Now the 2014 PN data are consistent with the 2000 data and the correction is sufficient for modelling.
		} 
		\label{Fig:Fe_Line_14}
	\end{figure}
	
	\section{Data reduction and spectral modelling}
	\label{Sec:Data_Reduction}
	NGC 1068 was observed twice in 2000 and four times in 2014 with XMM-Newton; the observation log is displayed in Table \ref{Table:Obs_Log}. We reduced the RGS and EPIC-PN data from each observation of NGC 1068 via the \texttt{SAS (v 17.0.0)} pipeline. The RGS data were reduced with \texttt{RGSPROC} and the EPIC-PN data were reduced using the \texttt{EPPROC} command. The 2000 observations were taken during the same XMM-Newton orbit so we stacked both the RGS and EPIC-PN spectra (separately) using \texttt{RGSCOMBINE} and \texttt{EPICSPECCOMBINE}, respectively. For the 2014 observations, separated by different time scales (see Table \ref{Table:Obs_Log}), there was little variability in the RGS spectra, so all four observations were combined. However, due to multiple flares (too close together to filter out the background noise) in the final 2014 observation (February 2015) we were only able to use and combine the first three (July and August 2014) EPIC-PN spectra; only the RGS data for the final 2014 observation were used (see Table \ref{Table:Obs_Log}). 
	
	At visual inspection, the \ion{Fe}{K$\alpha$} and higher energy emission lines in the 2014 EPIC-PN spectrum appear to be shifted with respect to the 2000 observation. The top panel of Fig. \ref{Fig:Fe_Line_14} compares the 2000 data (red crosses) with the 2014 data (blue crosses) for the \ion{Fe}{K$\alpha$} line (at around 6.4 keV), where there is an energy difference of $\Delta E = 50$ eV between the observed line peaks. This blueshift (and broadening; $\sigma(E) \sim 20$ eV relative to Chandra observations from 2012) of the \ion{Fe}{K$\alpha$} line was also found in NGC 5548, where the problem was attributed to uncertainties in the long-term charge transfer inefficiency and instrumental gain correction of the EPIC-PN CCDs, causing modest but significant variations \citep[][and references within]{Cappi2016}. Similar to \cite{Cappi2016}, we compared the 2014 PN spectrum to the 2000 and 2014 MOS spectra, as well as to some 2012 Chandra observations, and found that all \ion{Fe}{K$\alpha$} lines have energies consistent with 6.4 keV, except for the 2014 PN \ion{Fe}{K$\alpha$} line. Therefore, this blueshifted iron line is not a new and interesting result, but rather an artefact of the long life of the PN instrument.
	
	To compensate for this energy shift, we manually changed the gain correction of the PN instrument for the 2014 observations. Taking the event list files for each of the three observations in 2014 (the latter was not used due to background flaring), we multiplied the energy (\texttt{PI}) column by 0.993, before restacking the spectra. This procedure corrected the emission line energies in the 2014 PN data to be consistent with the 2000 data (see bottom panel of Fig. \ref{Fig:Fe_Line_14}).
	
\begin{figure*}
		\centering
		\begin{subfigure}{0.5\linewidth}
			\includegraphics[width=1\linewidth]{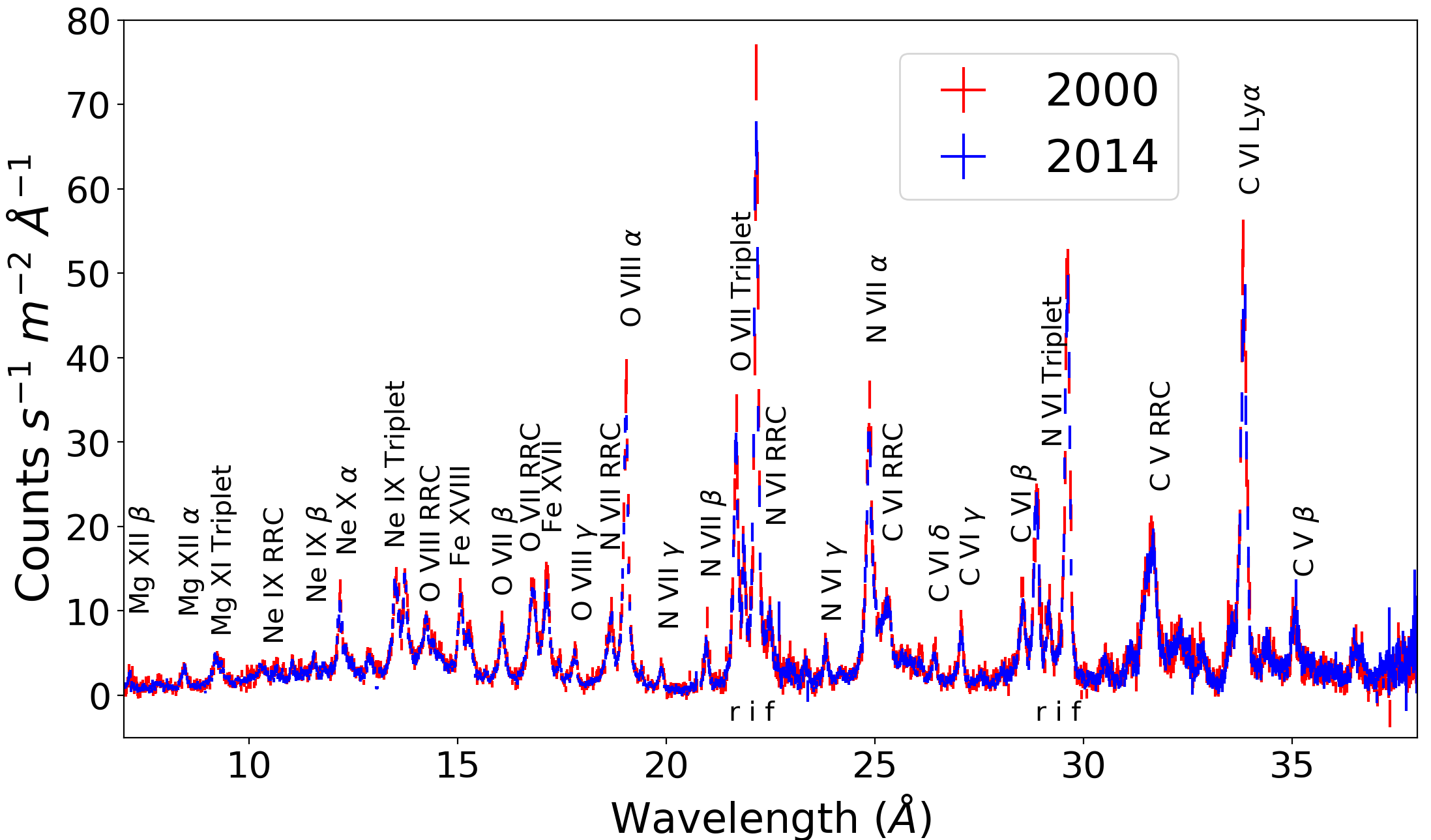}
		\end{subfigure}%
		\begin{subfigure}{0.5\linewidth}
			\includegraphics[width=1\linewidth]{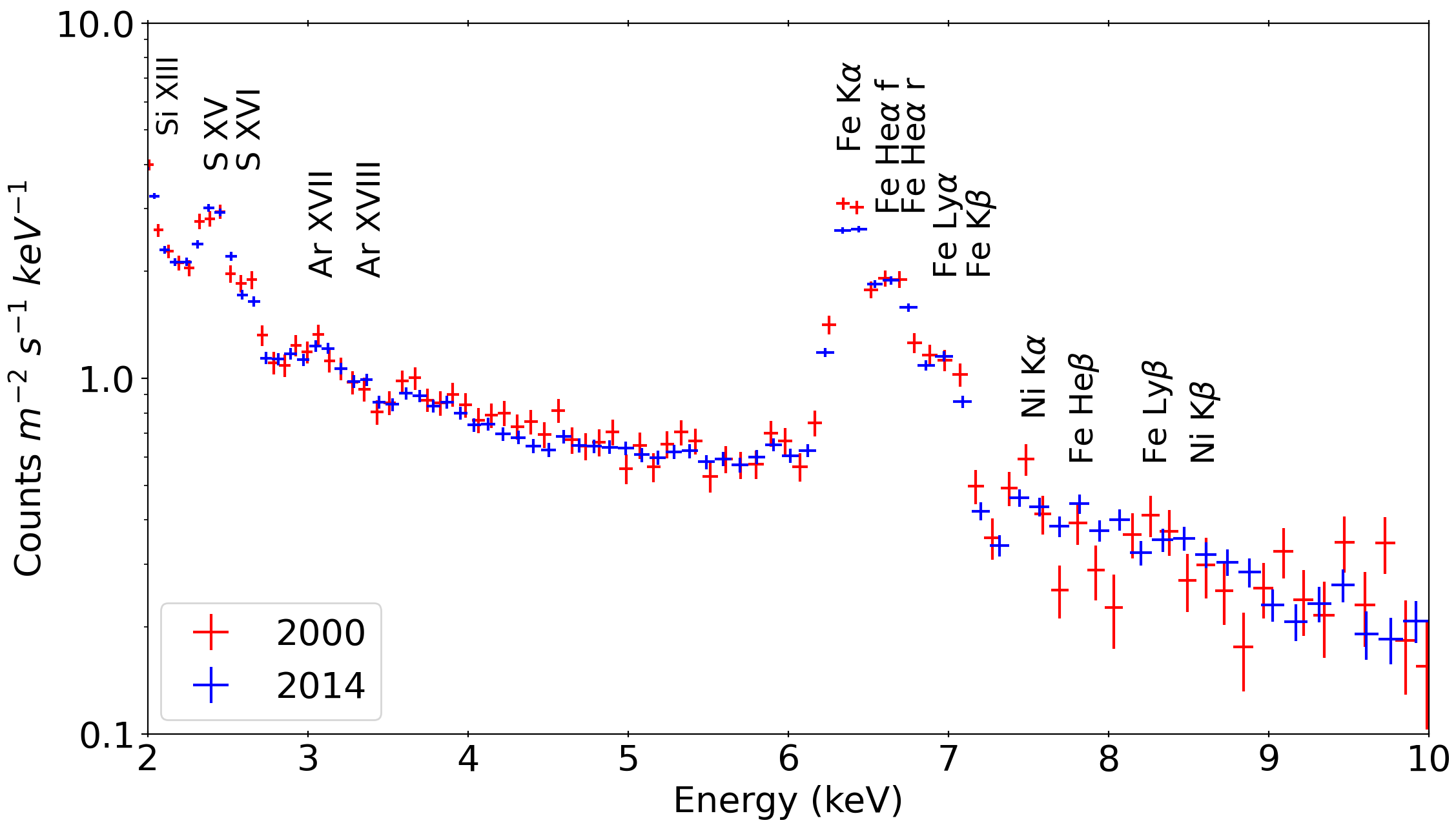}
		\end{subfigure}
		\caption{Comparing the RGS spectra (left) and EPIC-PN (right) spectra of NGC 1068 in 2000 (red) and 2014 (blue), in the observed reference frame. All spectra show a striking similarity in shape, emission features and flux, suggesting no change between the two epochs.}
		\label{Fig:Spec_Comparison}
	\end{figure*}

\begin{figure}
		\centering
		\includegraphics[width=1\linewidth]{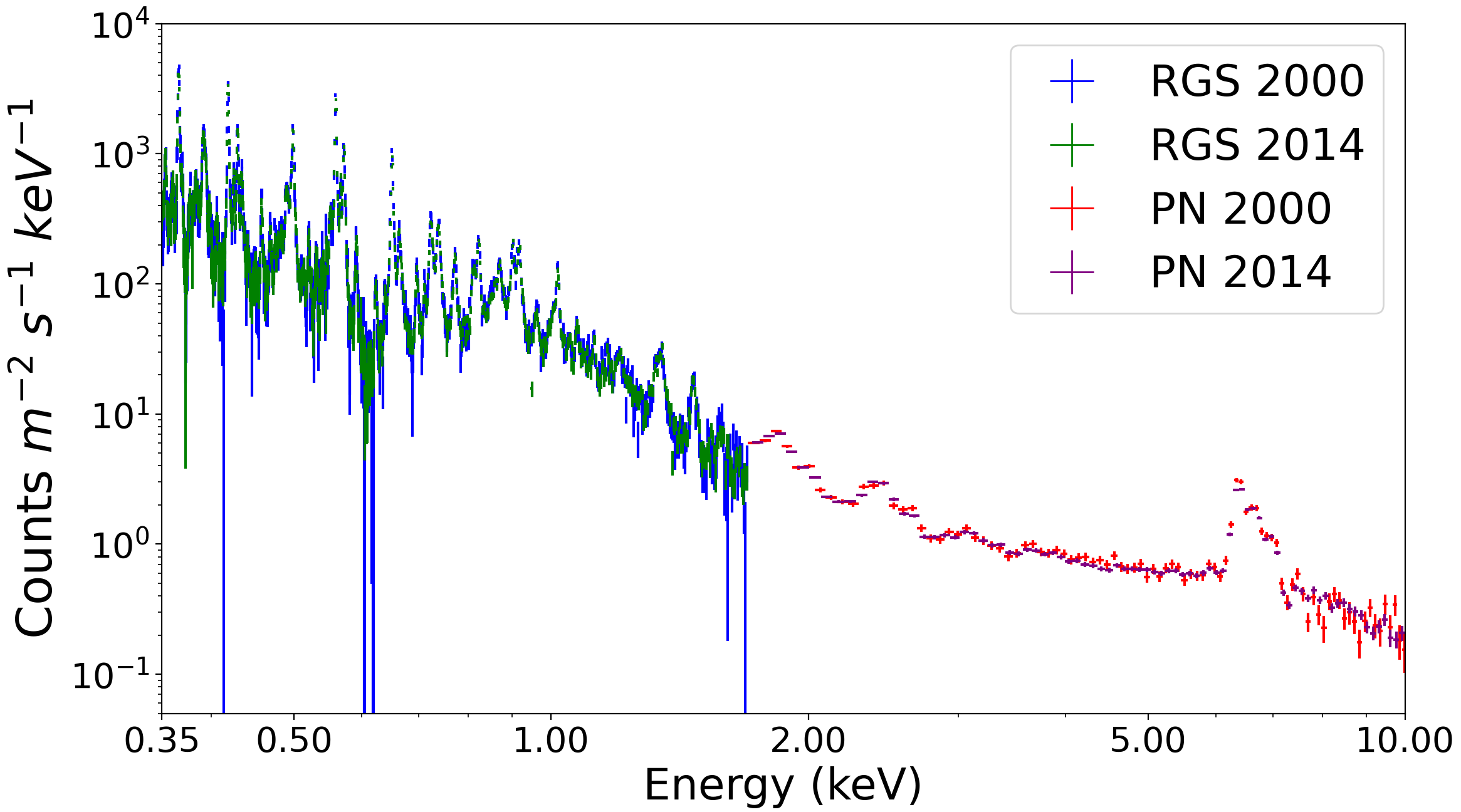}
		\caption{Combined RGS and PN spectra for the 2000 (blue and red) and 2014 (green and purple) observations, in the observed reference frame. We simultaneously fit both RGS and PN spectra for each epoch to model all emission features and observed continuum over the 0.35 - 10 keV energy range.}
		\label{Fig:Spec_PN_RGS}
	\end{figure}

Figure \ref{Fig:Spec_Comparison} displays the RGS (left) and PN (right) spectra of NGC 1068, in the observed reference frame, comparing both epochs, where the main emission lines are labelled with their respective ions. The spectra in Fig. \ref{Fig:Spec_Comparison} show very little to no variability between the two observations \citep[as noted by][]{Marinucci2016}. In order to conclusively determine the emission feature properties and their origins across the 0.35 - 10 keV energy range, we fitted both RGS and PN spectra simultaneously in each epoch (see Fig. \ref{Fig:Spec_PN_RGS}), thus allowing us to obtain better continuum constraints compared to using just RGS or PN data individually.	Although the EPIC-PN data could be used to search for subtle variations in the soft X-ray band, because of their higher count rate, we found possible pile-up in the 2000 data. There is a flux loss around 10 per cent in the 2000 PN spectra, and a spectral distortion around 2 per cent \citep[estimated as in][]{Jethwa2015}. However, as already noted by \cite{Matt2004}, these effects are negligible above 4 keV, while they become important at lower energies. As a result of this, and because there are almost no features in the two PN spectra below 2 keV, we ignored the PN spectra below 1.7 keV. This also reduced the weighted fitting bias as the PN continuum has higher flux counts at lower energies compared to the RGS spectrum. 
	
As the same continuum model goes through both RGS and PN data points, covering the whole X-ray band, it is acceptable to have no or little overlap between RGS and PN data. We wanted the high-resolution spectral features to be fitted well with RGS and the fitting not be affected by the statistically dominant PN data. This has been done in previous papers whereby the PN data were excluded in the RGS range \citep[see e.g.][]{Kaastra2014, Mehdipour2017}. Therefore, the RGS spectrum covers the energy range between 0.35 - 1.7 keV ($\sim 7 - 37$ \AA) and the PN spectrum accounts for the 1.7 - 10 keV range. We binned the EPIC-PN data using the optimal method as described by \cite{KB2016}, whereas the RGS data were binned by a factor of 3. This corresponds to an average bin size of 0.04 \AA, which still over-samples the RGS resolution element of $\sim 0.07$ \AA\ FWHM. This is shown in Figs. \ref{Fig:Spec_Comparison} (right side) and \ref{Fig:Spec_PN_RGS}.


	
	We modelled the combined spectra of NGC 1068 using \texttt{SPEX} \citep[\texttt{v 3.04.00;}][]{Kaastra1996}. In our model, we set the redshift of this AGN to z = 0.0038 \citep{Huchra1999} and assumed solar abundances from \cite{Lodders2009}. We use the Cash statistic \citep[C-statistic, hereafter;][]{Cash1979, Kaastra2017} for statistical significance, with errors at a 1$\sigma$ confidence.

	\section{Spectral energy distributions}
	\label{Sec:SED}
	\subsection{Ionising continuum}
	\label{Sec:ION_CONT}
	
	Due to the nature of NGC 1068, the LOS obscuration from the torus means we are unable to observe the central ionising source directly; this is the origin of the continuum that ionises the outflowing wind which we see either absorbing (Seyfert 1) or emitting (Seyfert 2) in the X-ray energy band \citep[see the cone model in e.g.][]{Kinkhabwala2002}. Therefore, as we cannot directly observe the ionising continuum, we had to assume a broadband continuum model, which we fixed throughout the spectral fitting. It is the shape of the ionising spectral energy distribution (SED) that is important, not the normalisation \citep{Mehdipour2016}, so it is not unreasonable to use a SED from a type 1 Seyfert galaxy. In this case, we adopted the SED of NGC 7469 derived by \cite{Mehdipour2018} as it is a bright and unobscured Seyfert 1 AGN, but we adjusted some of the parameters to match the observed values found for NGC 1068 in the literature \citep[e.g.][]{Bauer2015}. The SED shape of NGC 7469 is a good representation of a typical AGN SED when we do not know the actual ionising SED of NGC 1068, and is a step better than a simple power-law. \cite{Mehdipour2016} compared photoionisation results using the SEDs of NGC 5548 (both unobscured and obscured) and a power-law SED. We compared the SEDs and thermal stability S-curves of NGC 5548 \citep{Mehdipour2016} and NGC 7469 \citep{Mehdipour2018}, and found that within the X-ray ionisation region where we were probing ($\log \xi = 1 - 4$), the S-curves did not look significantly different. From this, we are confident that the differences between typical SEDs of the same type (like NGC 7469 and NGC 5548) would not significantly affect our results here. Therefore, NGC 7469 is a reasonable SED to use when modelling the ionising continuum of NGC 1068.

The three main components that contribute to the ionising SED are displayed in Table \ref{Table:Ion_SED}; they are as follows:
Firstly, a power-law (\texttt{POW}) was used to model the hard X-ray emission produced when optical/UV disk photons are upscattered in a hot electron corona. To this power-law we applied upper and lower cut-off energies. The lower cut-off energy was fixed at the disk temperature, adopted from NGC 7469 \citep{Mehdipour2018}, to stop the power-law energies becoming lower than the disk photon energies; the high energy cut-off was fixed at 500 keV \citep{Bauer2015}. 
The next SED component was a reflection component \citep[\texttt{REFL};][]{Magdziarz1995} which modelled the \ion{Fe}{K$\alpha$} line and Compton hump. The iron abundance ($A_{Fe} = 2.45$\footnote{Corrected for the abundances to \cite{Lodders2009}}), inclination ($\theta_{inc} = 60^\circ$) and photon index ($\Gamma = 2.1$, coupled to the photon index of \texttt{POW}) were fixed at the values found from previous observations of NGC 1068 \citep[e.g.][]{Matt2004, Pounds2006, Bauer2015}. Here we assumed that the X-rays we observed (after being reflected) were the same reflected X-rays that ionise the emitting plasma. This means that we had one \texttt{REFL} component that accounted for both ionising and observed X-ray reflection, so we fitted the normalisation parameter of \texttt{REFL} ($N_{refl}$) in the model.  This was the only fitted parameter in the SED, all other parameters were fixed to the values in Table \ref{Table:Ion_SED}.
Finally, a warm, optically thick medium has been found to best fit the soft X-ray excess and explain the UV/optical emission from the disk in Seyfert 1 AGN \citep[e.g.][]{Petrucci2013, Middei2018}, which we modelled with the Comptonisation component \citep[\texttt{COMT;}][]{Titarchuk1994}. All the values for \texttt{COMT} were adopted and fixed from the ionising SED in NGC 7469 \citep{Mehdipour2018}.

	\subsection{Observed continuum}
	\label{Sec:Obs:Cont}
	The X-ray spectrum we do see is the result of the ionising continuum reflecting off the circumnucleur material within or surrounding the nucleus of NGC 1068 \citep[e.g. scattering off the dusty torus, or off free electrons;][]{Antonucci1985}, interacting with and ionising the outflowing plasma wind. This observed continuum was modelled with a reflection component (same \texttt{REFL} as the ionising SED; see above) for the hard X-rays, with the addition of a simple power-law component for the soft X-ray band, for completeness. As we are only interested in the photoionised emission lines within NGC 1068, an ad hoc continuum, in the form of a simple observed power-law plus reflection, is justified.
	
	
	\begin{table}
		\centering
		\caption{Parameter values (fixed) for the ionising SED initially adopted from the broad band ionising continuum of NGC 7469 \citep[][see Sect. \ref{Sec:ION_CONT} for details]{Mehdipour2018}. To modify the SED to match with NGC 1068, some of the parameters were changed to observational results found in the literature, shown in the reference column. We assumed the ionising reflection was the same as the observed reflection, so we fitted only the normalisation of the \texttt{REFL} component.}
		\label{Table:Ion_SED}
		\begin{tabular}{c | c c | c}
			\hline \hline
			Component & Parameter & Value & Reference \\
			\hline
			\multirow{2}{*}{\texttt{POW}\ \ \tablefootmark{a}} & Norm\ \ \tablefootmark{b} & $7.1 \times 10^{51}$& 1 \\
			& $\Gamma$ & 2.1 & 2, 3, 4 \\ 
			\hline
			\multirow{4}{*}{\texttt{COMT}} & Norm\ \ \tablefootmark{c}& $5.7 \times 10^{55}$ & \multirow{4}{*}{1} \\
			& $T_{disk}$ (eV) & 1.4 & \\
			& $T_e$ (keV)& 0.14& \\
			& $\tau$ & 21& \\
			\hline
			\multirow{6}{*}{\texttt{REFL}} & $\Gamma$ \ \ \tablefootmark{d} & 2.1& 2, 3, 4 \\
			& $A_{Fe}$ & 2.45\ \ \tablefootmark{e}  & 2, 3 \\
			& $E_c$ (keV) & 500 & 3 \\
			& $\theta_{inc}$ & 60$^\circ$ & 3 \\
			& Reflection scale & 0.5  & -  \\
			& $\sigma_v$ (km s\textsuperscript{-1})& $\sim 3000$ & 1 \\
			\hline 
			\hline
		\end{tabular}
		\tablefoot{
			\tablefoottext{a}{We applied an upper and lower cut-off energy to the power-law: the lower cut-off energy ($E_L$) is equal to the disk temperature ($T_{disk}$) and the upper cut-off energy is $E_C = 500$ keV \citep{Bauer2015}.} The units for the normalisations are:
			\tablefoottext{b}{ph s\textsuperscript{-1} keV\textsuperscript{-1} at 1 keV; and} 
			\tablefoottext{c}{ph s\textsuperscript{-1} keV\textsuperscript{-1}.}
			\tablefoottext{d}{Coupled to \texttt{POW}.} \tablefoottext{e}{This value has been adjusted from \cite{Matt2004, Pounds2006}, who measured $A_{Fe} = 2.4$, to the abundance of \cite{Lodders2009}.}
		}
		\tablebib{(1)~\citet{Mehdipour2018}; (2)~\citet{Matt2004}; (3)~\citet{Bauer2015}; (4)~\citet{Pounds2006}.}
	\end{table}

	\section{Data analysis}
	\label{Sec:Data_Analysis}
	
	In both 2000 and 2014 spectra, we modelled the observed continuum by fixing both the power-law (\texttt{POW}) and reflection (\texttt{REFL}) parameters to the values in Table \ref{Table:Ion_SED}. However, we did fit the normalisation parameters of both \texttt{POW} ($N_{pow}$) and \texttt{REFL} ($N_{refl}$; same as the ionising reflection component), and the photon index ($\Gamma$) for the power-law, which we initially set to 2.1; $\Gamma$ in \texttt{REFL} was coupled and fixed at this value \citep{Matt2004, Pounds2006, Bauer2015}. We also coupled a \texttt{VGAU} component to \texttt{REFL} to obtain a measurement for the broadness ($\sigma_v$) of the \ion{Fe}{K$\alpha$} line, measured as a velocity. 
	
	The Galaxy absorption was modelled with the \texttt{HOT} component, setting the total hydrogen column density to $N_H^{Gal} = 3.34 \times 10^{24}$ m\textsuperscript{-2}, made up from both \ion{H}{I} \citep{Kalberla2005} and H\textsubscript{2} \citep{Wakker2006}. For a neutral Galactic gas, we fixed the temperature and turbulent velocity to $T_{Gal} = 0.5$ eV and $v_{turb}^{Gal} = 5.62 \pm 3.19$ km s\textsuperscript{-1} \citep[for details of obtaining this turbulent velocity result, see the Appendix in][]{Grafton-Waters2020}, respectively.  All fitted components were redshifted \citep[z = 0.0038;][]{Huchra1999} and then absorbed by the Galaxy medium in this model. 
	
	We modelled the photoionised emission lines in NGC 1068 using the sophisticated and self-consistent photoionisation model \texttt{PION} \citep{Mehdipour2016} in \texttt{SPEX}. \texttt{PION} uses the full broadband SED continuum \citep[in this case adopted from NGC 7469;][]{Mehdipour2018} to calculate the emission spectrum and the ionisation/thermal balance of the plasma, simultaneously. To fit the emission lines in both epochs, we fitted one component at a time, fixing the previous component before adding the next, until no further improvement on the overall fit statistic was achieved. In all \texttt{PION} components, we fitted the column density ($N_H$), ionisation parameter ($\xi$), covering fraction ($C_{cov}$), and outflow and turbulent velocities ($v_{out}$ and $v_{turb}$, respectively). 

	After fitting the photoionised emission (\texttt{PION}) components, we still found residuals in some of the emission features (mostly in the region between 14 - 18 \AA). Although the large consensus regarding CIE plasma is that it is not the dominant source for X-ray emission in the nucleus of NGC 1068 \citep[e.g.][]{Young2001,Kinkhabwala2002,Ogle2003}, there is evidence of a star burst region \citep[e.g.][]{Garcia-Burillo2014}. Therefore, we introduced a \texttt{CIE} component to the model to see if this could explain the residuals in the spectra, fitting the electron temperature ($T_{e}$), emission measure (EM), and outflow ($v_{out}$) and turbulent ($v_{t}$) velocities. This was not to account for the lower order resonance lines \citep[expected to be produced via photoexcitation processes;][]{Kinkhabwala2002}, as \texttt{PION} takes these into consideration, but rather to fit the residuals in the emission spectrum between 14 and 18 \AA. Figure \ref{Figure:PION_CIE} shows that some of the emission lines in the 2000 RGS spectrum could not be explained by PIE plasma (red line), but were accounted for when we fitted for CIE plasma (orange line). In NGC 7469, \cite{Behar2017} found residuals in the RGS spectrum at 15.3 and 17.4 \AA\ (\ion{Fe}{XVII}) which they interpreted as emission in CIE plasma from the star burst region. Therefore, we included a \texttt{CIE} component in our modelling.

	The elemental abundances fitted in this model were for C, N, O, Ne, Mg, Si, S, Ar, Ca and Ni, all with respect to Fe \cite[the abundance of Fe was found to be $A_{Fe} = 2.45$\footnote{After adjusting for the abundances in \cite{Lodders2009}.}, relative to Solar, in NGC 1068;][]{Matt2004, Bauer2015} as Fe emission lines were found in both RGS and PN spectra (oxygen is the strongest feature only in the RGS energy band). However, as C, N, O, Ne and Mg were only found in the RGS energy range, and Si, S, Ar, Ca and Ni were only present in the PN spectrum, we split the abundances up into two \texttt{PION} components when fitting. The first \texttt{PION} component fitted the high energy emission lines of the five higher Z elements and the second \texttt{PION} component (RGS band) accounted for the five lower Z elements. The remaining \texttt{PION} and \texttt{CIE} components had their abundance values coupled to these two \texttt{PION} components. This meant that the model fitted the abundances for all the lines from each component, but with each element free only once. In order to do this, we assumed that the chemical enrichment is the same in all emitting plasma regions (photoionised or collisionally ionised) throughout the AGN nucleus and star burst region of NGC 1068.

	
	\section{Spectral results}
	\label{Sec:Results}
	In Fig. \ref{Fig:Spec_PN_RGS}, the combined RGS and PN spectra show many complex features. This is a result of a mixture of many emission lines, some of which are blended, and an obscured continuum that is not easily modelled. This meant that acquiring a statistically significant best fit, represented by the C-statistic, was very challenging. Therefore, in both epochs, we used the change in C-statistic ($\Delta C$) to signify the improvement on the best fit from each fitted component. Despite this however, $\Delta C$ could still be very large due to the complexity of the spectrum and its statistical quality. 
	
	\subsection{2000 best fit}
	\label{Sec:Results_2000}
	The initial fitting of the observed continuum (\texttt{POW} and \texttt{REFL}) gave a C-statistic of C = 58486 (for 1075 degrees of freedom; hereafter d.o.f). Throughout the emission component fitting (\texttt{PION} and \texttt{CIE}), the continuum parameters were free to vary. The change in C-statistic was with respect to the previous component, which we fixed when introducing the next component, unless stated otherwise. 
	
	The first \texttt{PION} emission component (EM1) was required to fit the high energy emission lines in the PN spectrum (E > 6.4 keV). This highly ionised ($\log \xi = 3.91$) component improved the fit with a fairly modest $\Delta C = 2200$ compared to the continuum. The second \texttt{PION} component (EM2; $\log \xi = 0.63$) accounted for many of the narrow emission lines in the RGS spectrum, improving the quality of the fit by $\Delta C = 35600$. When we fitted both EM1 and EM2 together, the fit improved by $\Delta C = 270$. A third \texttt{PION} component (EM3; $\log \xi = 1.84$) fitted lines in both RGS and PN spectra, improving the fit by $\Delta C = 4700$. Again, we then freed all (three) \texttt{PION} components, decreasing the statistic by $\Delta C = 1600$. A fourth component (EM4; $\log \xi = 3.02$) further improved the fit, albeit not as significantly as the other three components, by $\Delta C = 220$. All four \texttt{PION} components were then fitted together, improving the model significantly by a further $\Delta C = 240$.
	
	The \texttt{CIE} component (CI1) improved the fit by $\Delta C = 1440$ (after fixing all the \texttt{PION} components), accounting for all the emission lines not produced by photoionised plasma, with large enough statistical significance. We then freed and refitted all four photoionised \texttt{PION} components with the \texttt{CIE} component to obtain a change in the C-statistic of $\Delta C = 840$.

	Finally, we fitted the abundances of the ten elements in the spectrum, with respect to iron. Firstly, we fixed all the parameters in the \texttt{PION} and \texttt{CIE} components, in order not to fit too many parameters at the same time when adding some new free parameters. Otherwise, this could cause a degenerate result for the parameter values for the same $\Delta C$. We freed and refitted the emission components after we accounted for the abundances. Therefore, we fitted the abundances of the elements in EM2 for the RGS spectrum (C, N, O, Ne and Mg) followed by the high energy elements (Si, S, Ar, Ca and Ni) in EM1. The change in C-statistic for these abundances were $\Delta C = 4900$ and $\Delta C = 330$, respectively. From this, we slowly fitted the emission components, firstly EM1 and EM2 ($\Delta C = 220$) followed by EM3 and EM4 ($\Delta C = 500$) and then CI1 ($\Delta C = 500$), with all abundances left free throughout these last few steps. Here, all parameters were fitted to obtain a new global C-statistic minimum of C = 4970 for 1040 d.o.f.
	
	Despite obtaining a relatively good fit so far, there was still an emission feature not accounted for. This line, at around 7.4 keV, belonging to the neutral \ion{Ni}{K$\alpha$}, comes from the X-ray reflection off neutral material similarly to the \ion{Fe}{K$\alpha$} line. As \texttt{REFL} in \texttt{SPEX} only accounts for neutral Fe, we fitted this line with a Gaussian (\texttt{GAUS}; for simplicity), freeing the normalisation, energy and line broadening, which improved the fit by $\Delta C = 130$. A similar approach was taken by \cite{Semena2019} for NGC 5643. Table \ref{Table:Ni_Results} shows the best fit parameter values of the \ion{Ni}{K$\alpha$} line.


All the parameters were then refitted. The best fit for our model, simultaneously applied to both the RGS and PN spectra, was C = 4530 for 1037 d.o.f. The final best fit model, including more \texttt{GAUS} components (see Sect. \ref{Sec:RGS_PN_Sep}), over imposed on the spectrum is shown in Fig. \ref{Fig:RGS_PN_Best_Fit_2000} and the best fit parameter values for the continuum, \texttt{PION}, \texttt{CIE}, and abundances are displayed in Tables \ref{Table:Cont_Results} - \ref{Table:Abund:Results}, respectively. All parameter errors are obtained with all fitted parameters free.

	\begin{figure}
		\centering
		\includegraphics[width=1\linewidth]{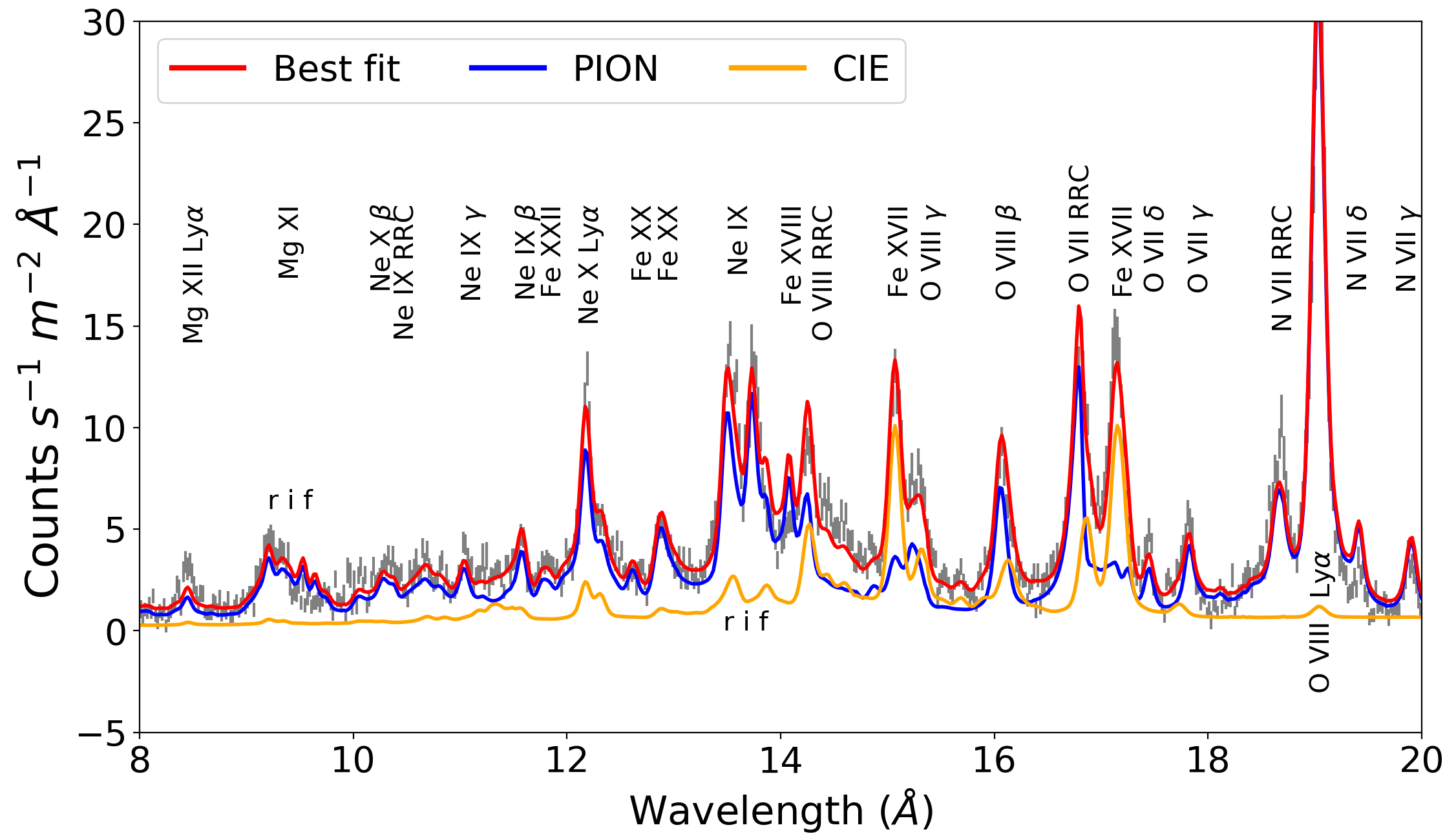}
		\caption{RGS spectrum (grey points) between 8 and 20 \AA\ from the 2000 observation of NGC 1068. The red line shows the best fit model and the blue line displays the photoionisation model made up of four different \texttt{PION} components, fitting the majority of the emission lines in both the RGS and PN spectra, except for the features at 15 and 17 \AA. We therefore introduced a collisionally ionised component (\texttt{CIE}; orange line) to account for these emission lines from \ion{Fe}{XVII}.}
		\label{Figure:PION_CIE}
	\end{figure}
	
	\begin{figure*}
		\centering
		\begin{subfigure}{0.5\linewidth}
			\includegraphics[width=1\linewidth]{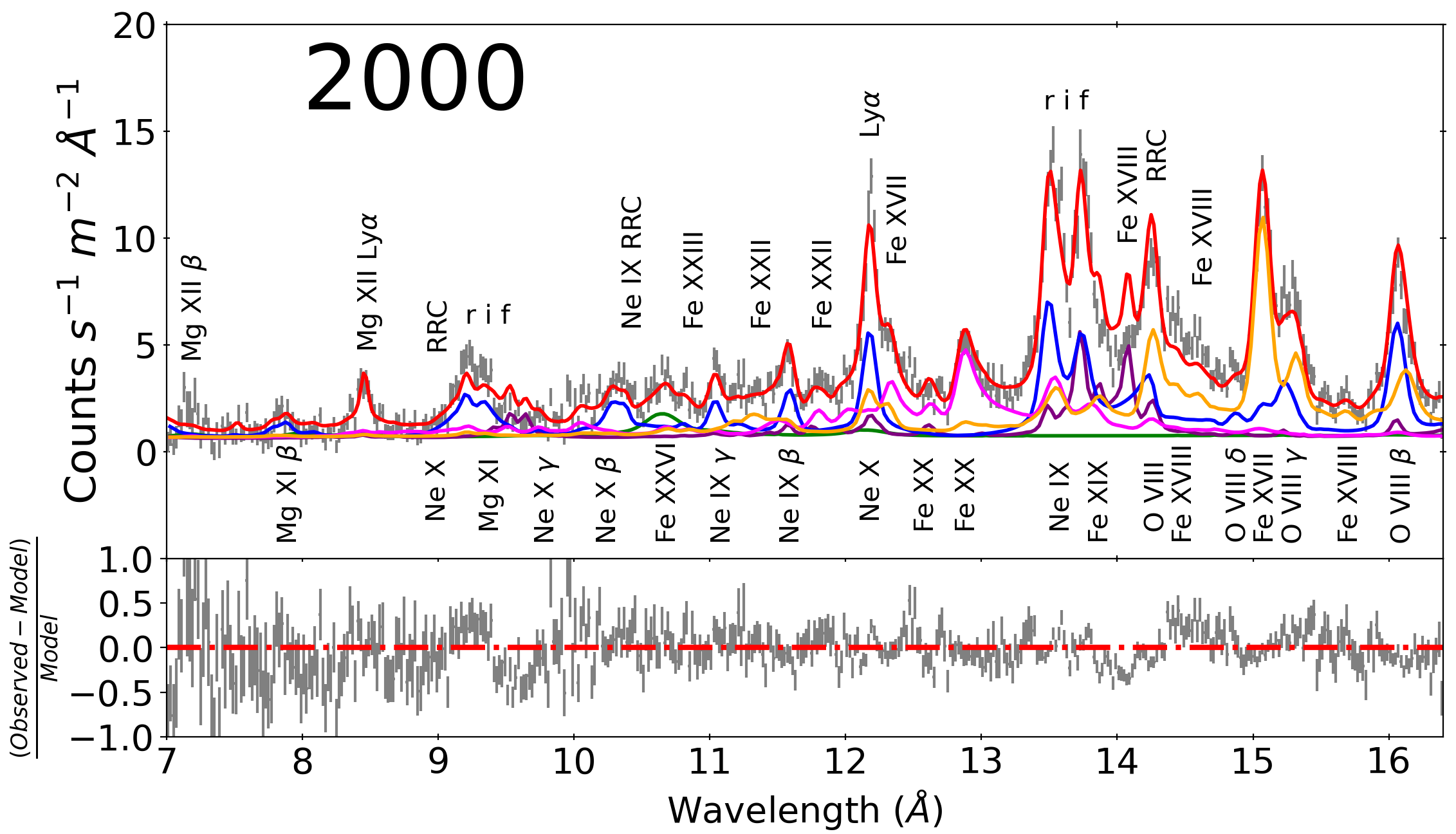}
		\end{subfigure}%
		\begin{subfigure}{0.5\linewidth}
			\includegraphics[width=1\linewidth]{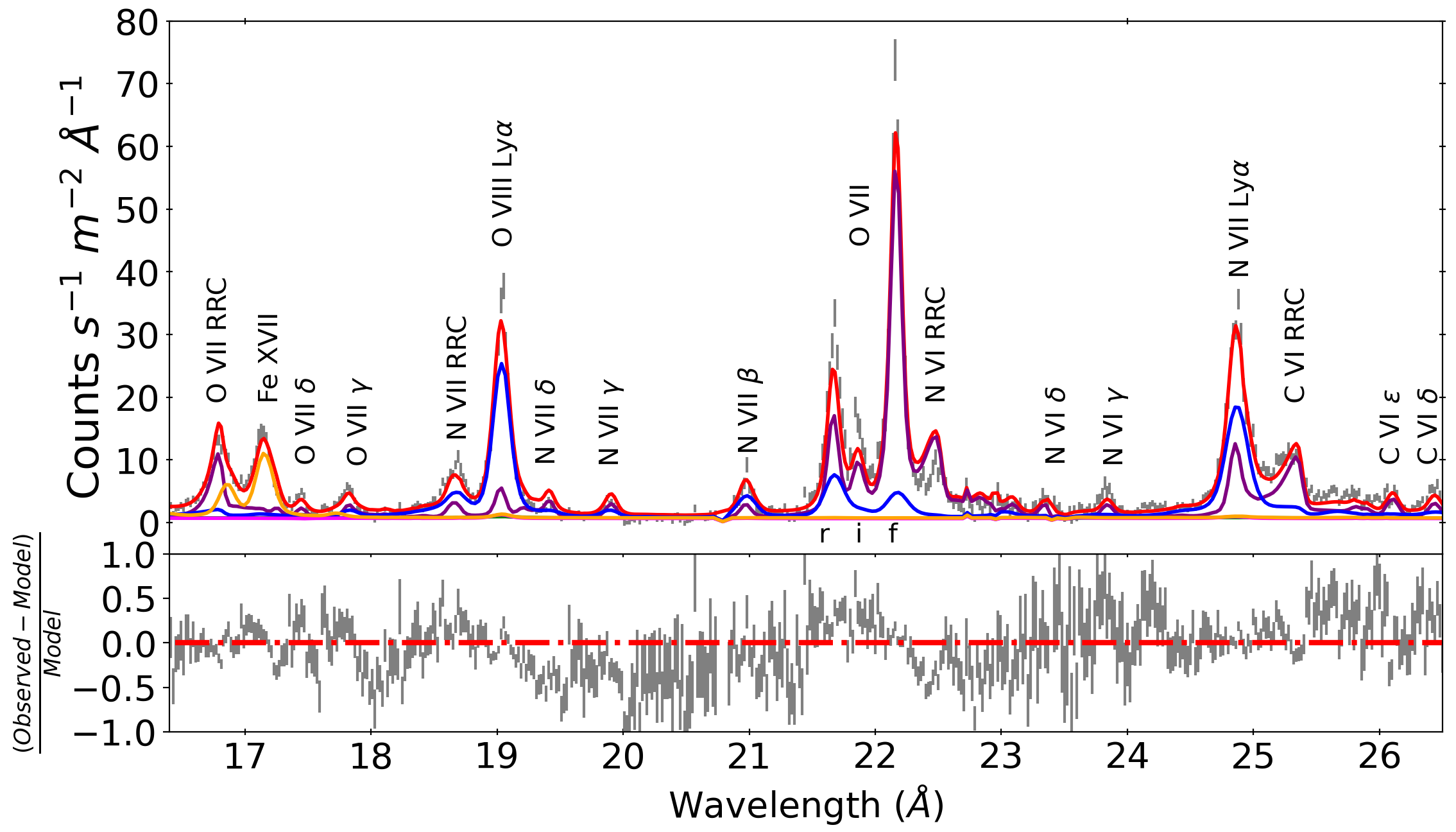}
		\end{subfigure}
		\begin{subfigure}{0.5\linewidth}
			\includegraphics[width=1\linewidth]{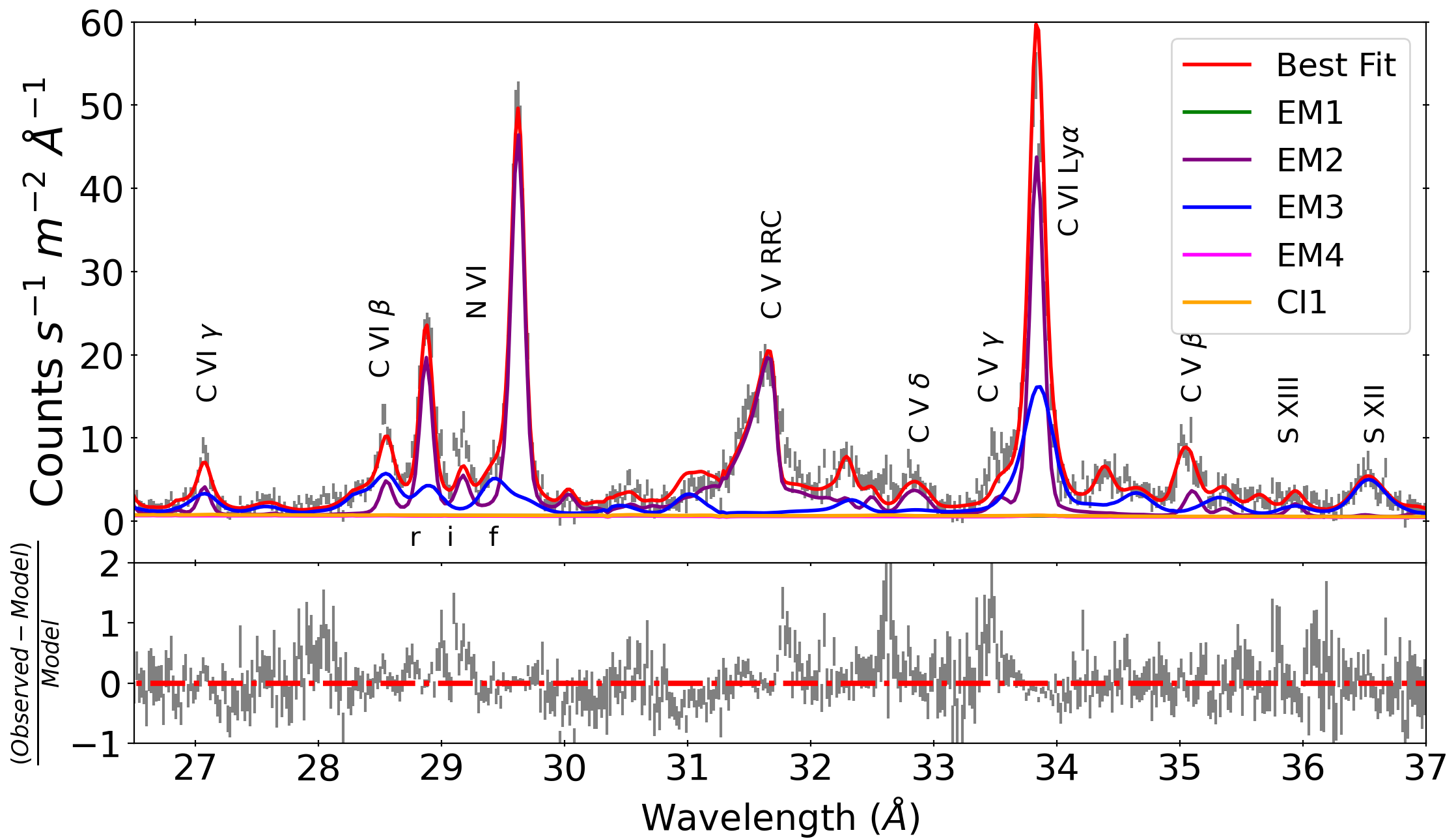}
		\end{subfigure}%
		\begin{subfigure}{0.5\linewidth}
			\includegraphics[width=1\linewidth]{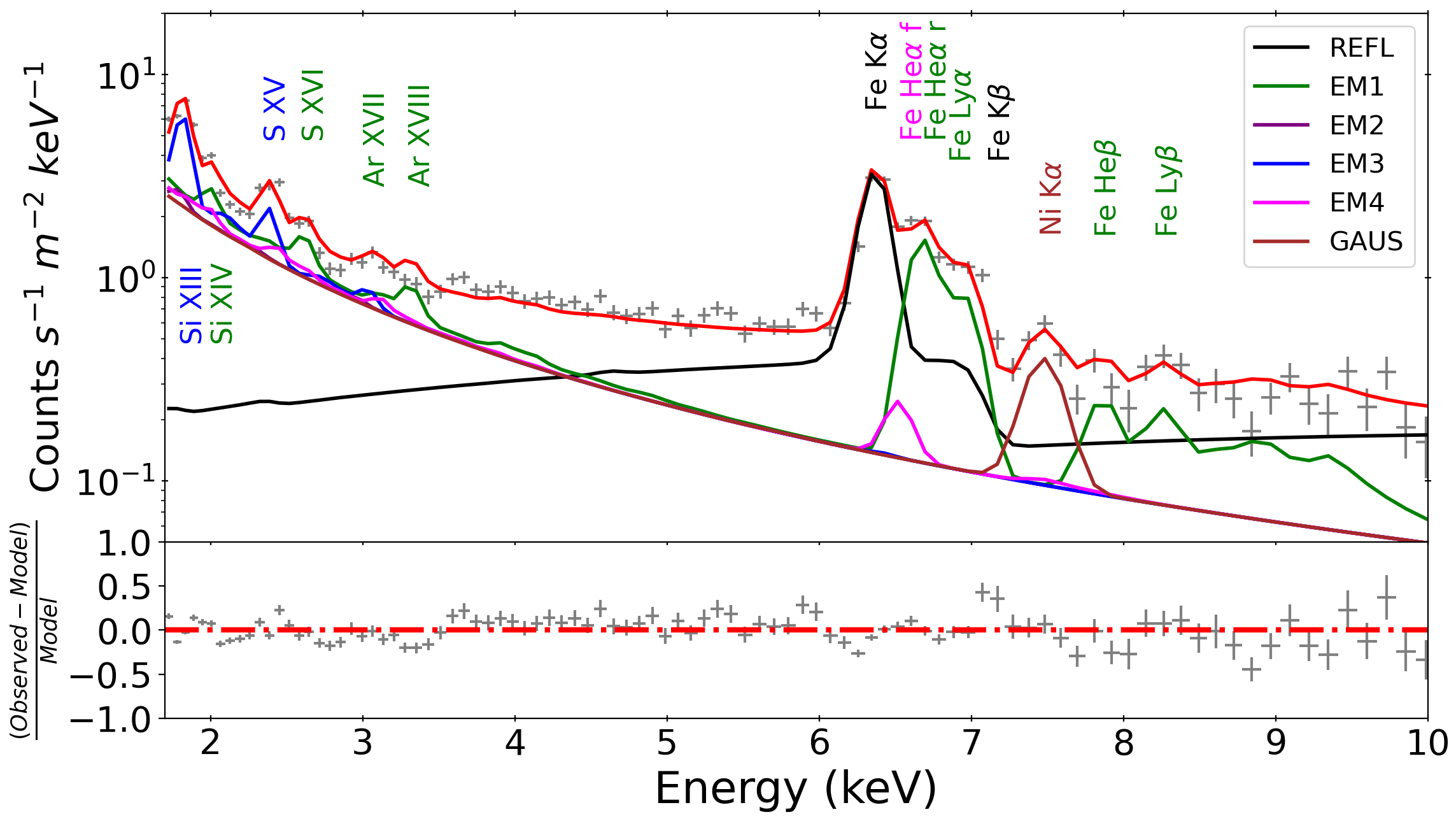}
		\end{subfigure}
		\caption{Best fit model to the 2000 spectrum of NGC 1068. We plot the soft X-ray (RGS) band in wavelength units and the hard X-ray (EPIC-PN) band in energy units for display purposes. The red line shows the best fit to the data points (grey crosses) and the other coloured lines (labelled in the legends) represent each component in the model. The bottom panels display the residuals between the best fit model and the observed data points.}
		\label{Fig:RGS_PN_Best_Fit_2000}
	\end{figure*}

	\begin{figure*}
		\centering
		\begin{subfigure}{0.5\linewidth}
			\includegraphics[width=1\linewidth]{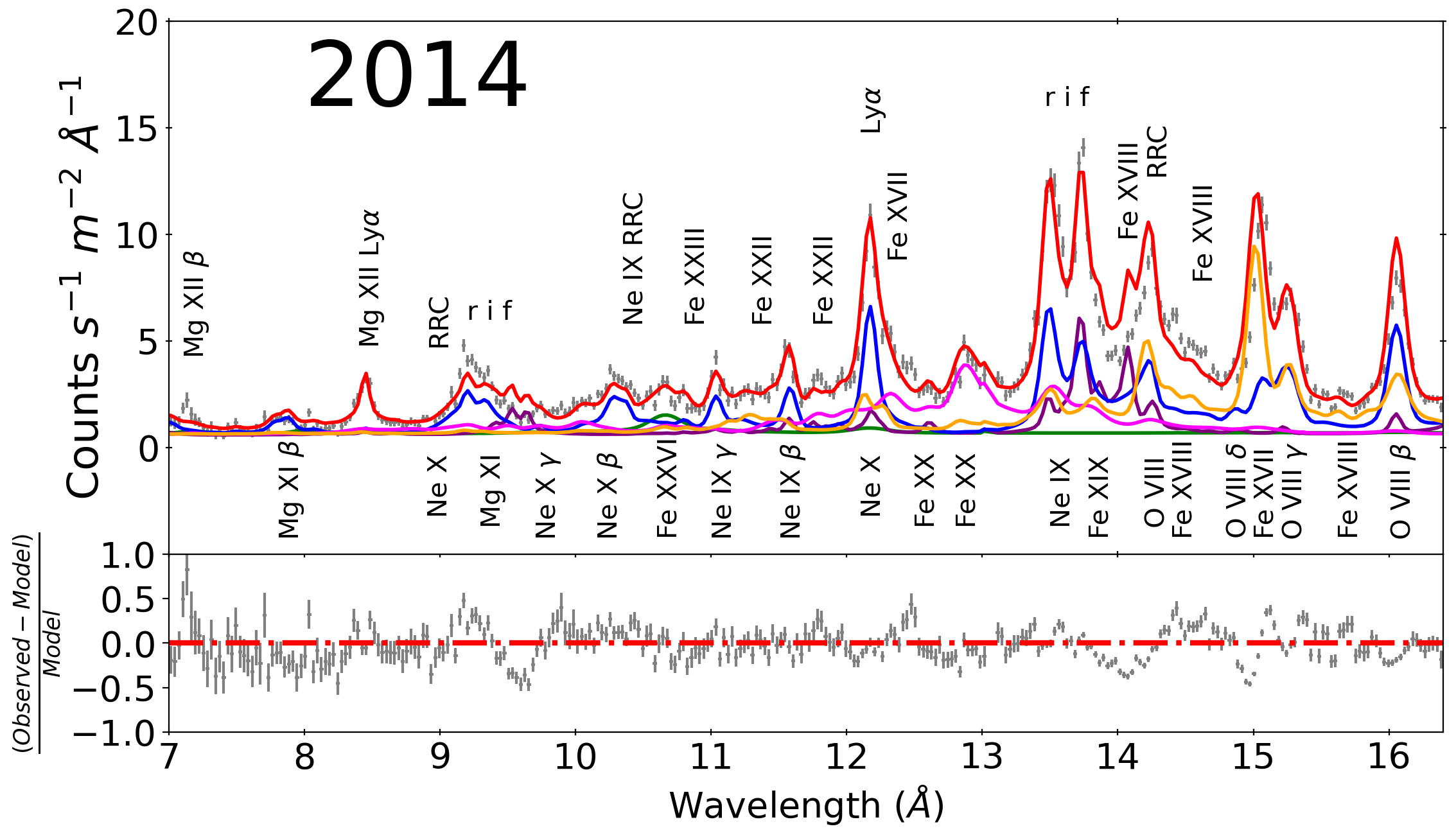}
		\end{subfigure}%
		\begin{subfigure}{0.5\linewidth}
			\includegraphics[width=1\linewidth]{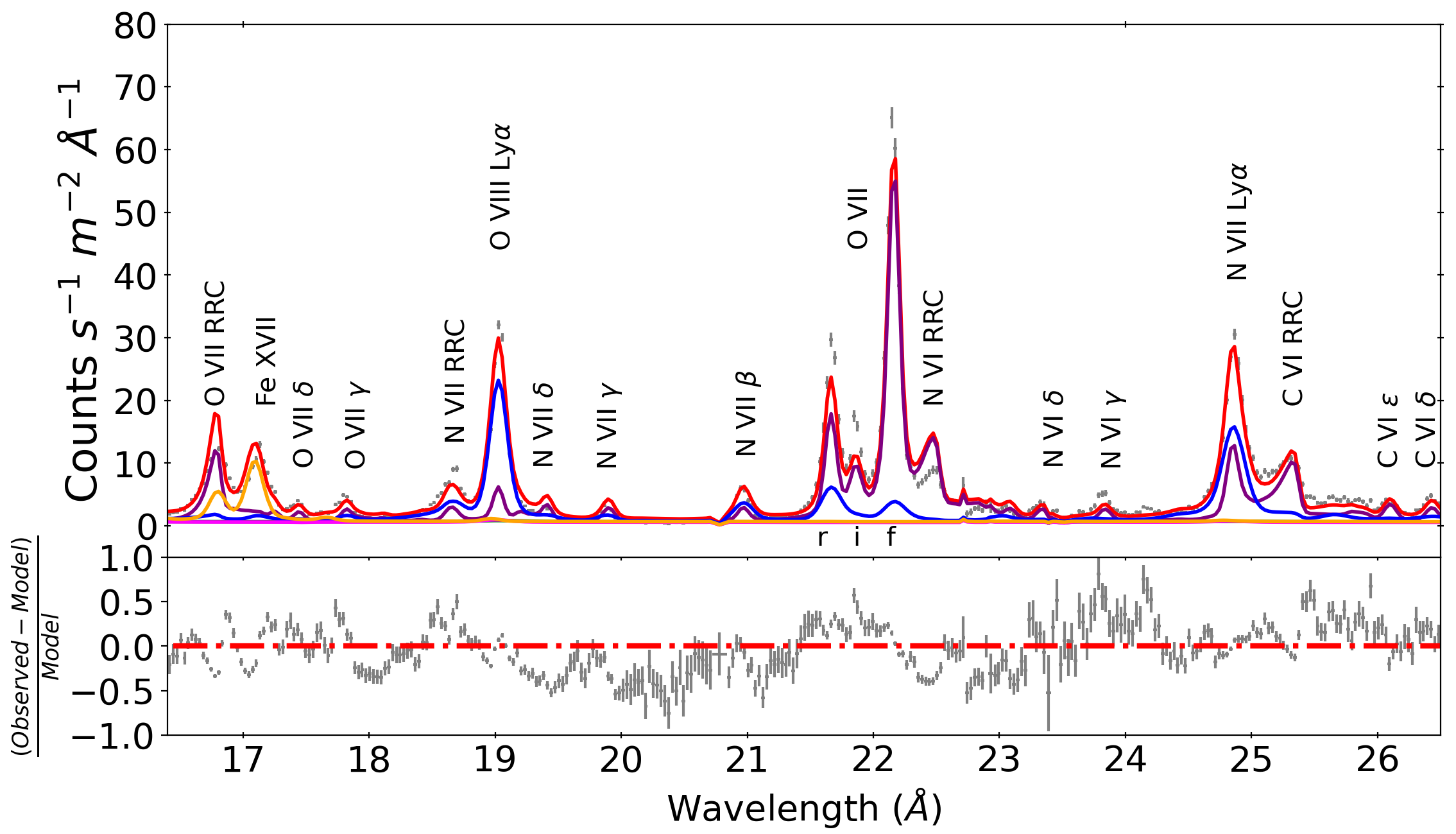}
		\end{subfigure}
		\begin{subfigure}{0.5\linewidth}
			\includegraphics[width=1\linewidth]{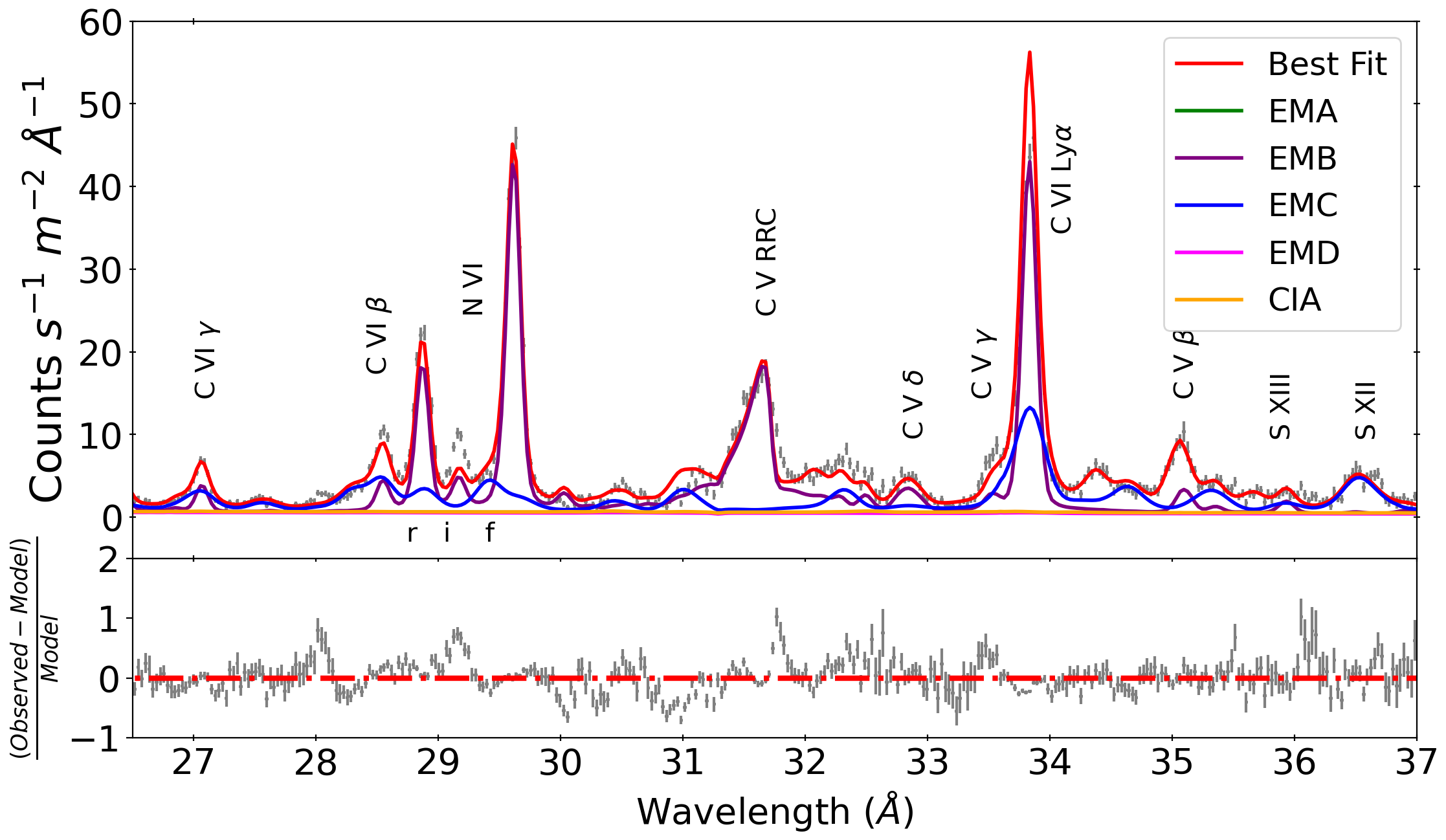}
		\end{subfigure}%
		\begin{subfigure}{0.5\linewidth}
			\includegraphics[width=1\linewidth]{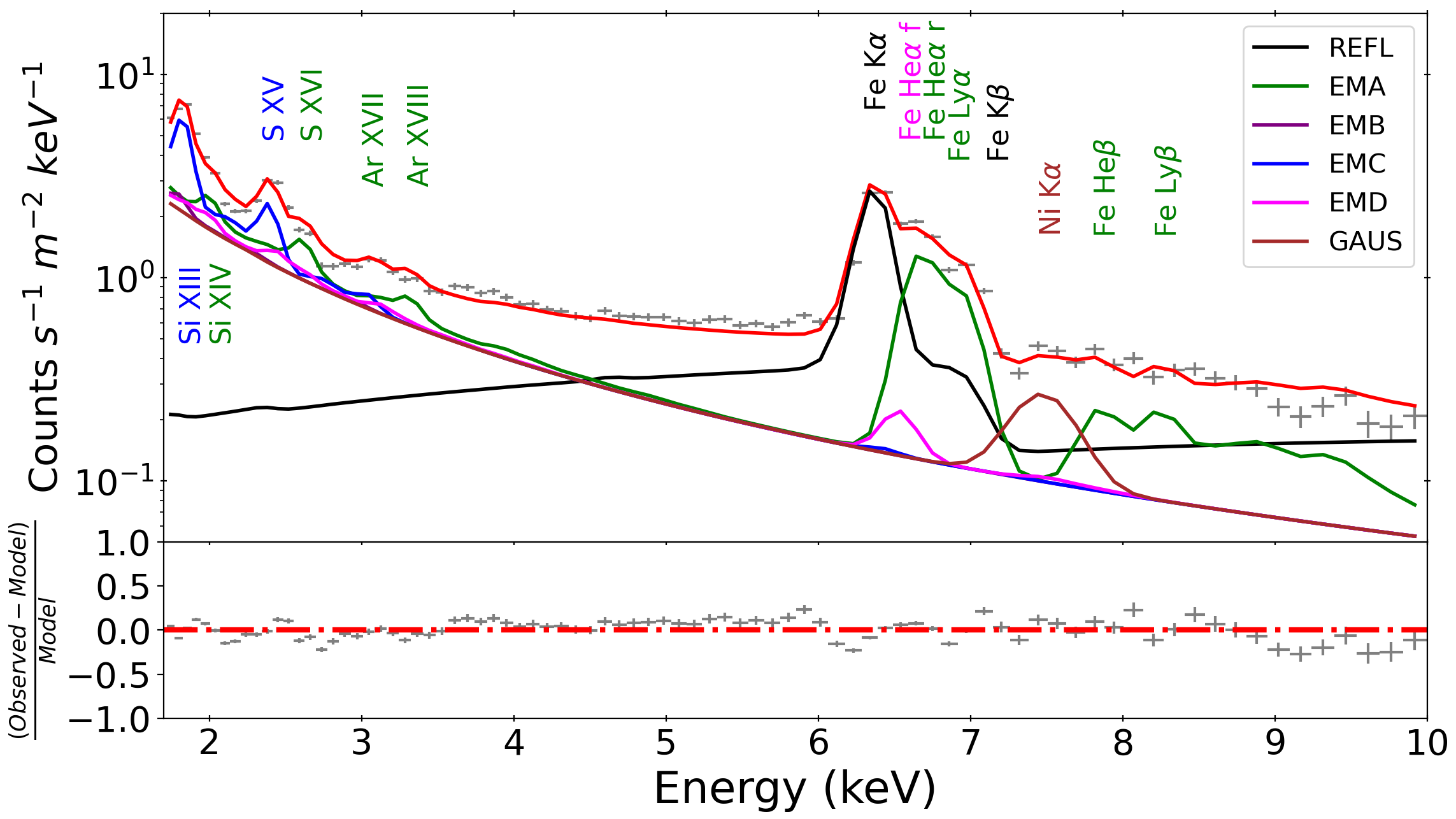}
		\end{subfigure}
		\caption{Best fit model to the 2014 spectrum of NGC 1068. We plot the soft X-ray (RGS) band in wavelength units and the hard X-ray (EPIC-PN) band in energy units for display purposes. The red line shows the best fit to the data points (grey crosses) and the other coloured lines (labelled in the legends) represent each component in the model. The bottom panels display the residuals between the best fit model and the observed data points.}
		\label{Fig:RGS_PN_Best_Fit_2014}
	\end{figure*}
	
	\begin{table}
		\centering
		\caption{Observed continuum best fit parameter values for the 2000 (top) and 2014 (bottom) observations; $\sigma_v$ corresponds to the width of the \ion{Fe}{K$\alpha$} line. }
		\label{Table:Cont_Results}
		\begin{tabular}{c | c c c}
			\hline
			Obs. & Component & Parameter & Value \Tstrut\Bstrut \\
			\hline
			
			\hline
			\multirow{5}{*}{2000} & \multirow{2}{*}{\texttt{POW}} & Norm\ \ \tablefootmark{a} &  $2.27^{+0.12}_{-0.05}$\Tstrut\Bstrut \\
			& & $\Gamma$ & $2.24 \pm 0.02$ \Tstrut\Bstrut \\
			& \multirow{2}{*}{\texttt{REFL}} & Norm\ \ \tablefootmark{b}  & $9.47^{+0.23}_{-0.24}$ \Tstrut\Bstrut \\
			& & $\sigma_v$ (km s\textsuperscript{-1}) &  $2120^{+270}_{-360}$\Tstrut\Bstrut \\
			\hline
			\multirow{5}{*}{2014} & \multirow{2}{*}{\texttt{POW}} & Norm\ \ \tablefootmark{a} & $2.06^{+0.03}_{-0.08}$ \Tstrut\Bstrut \\
			& & $\Gamma$ & $2.18^{+0.02}_{-0.04}$\Tstrut\Bstrut \\
			& \multirow{2}{*}{\texttt{REFL}} & Norm\ \ \tablefootmark{b} &  $8.86^{+0.15}_{-0.26}$ \Tstrut\Bstrut \\
			& & $\sigma_v$ (km s\textsuperscript{-1}) & $< 1800$ \Tstrut\Bstrut \\
			\hline
		\end{tabular}
		\tablefoot{
			\tablefoottext{a}{$\times 10^{49}$ ph s\textsuperscript{-1} keV\textsuperscript{-1} at 1 keV;}
			\tablefoottext{b}{$\times 10^{50}$ ph s\textsuperscript{-1} keV\textsuperscript{-1} at 1 keV.}
		}
	\end{table}

	\begin{table*}
		\centering
		\caption{Best fit parameter values for the \texttt{PION} components fitted to the 2000 (top) and 2014 (bottom) RGS and EPIC-PN combined spectra. }
		\label{Table:PION_Results}
		\begin{tabular}{c | c | c c c c c c| c}
			\hline
			\multirow{2}{*}{Obs.} &\texttt{PION}& $N_H$ & $\log \xi$ & $v_{turb}$ & $v_{out}$ & $C_{cov} =$ & EM&\multirow{2}{*}{$\Delta C$} \\
			& Component & ($10^{25}$ m\textsuperscript{-2}) & ($10^{-9}$ W m) & (km s\textsuperscript{-1}) & (km s\textsuperscript{-1}) & $ \Omega / 4\pi$ & $10^{69}$ (m\textsuperscript{-3}) &\\
			\hline
			\hline
			\multirow{5}{*}{2000} & EM1 & $50^{+1}_{-2}$ & $3.91 \pm 0.02$ & $2910^{+280}_{-260} $ & $-610^{+200}_{-330}$ & $0.19 \pm 0.01$ & $2.7^{+0.1}_{-0.2}$ & 2200\Tstrut\Bstrut \\
			& EM2 & $41 \pm 2 $ & $0.63 \pm 0.01 $ & $370 \pm 10$ & $-260 \pm 10 $ & $0.03 \pm 0.01$ & $670^{+270}_{-240}$ & 35600\Tstrut\Bstrut \\
			& EM3 & $22 \pm 1$ & $1.84 \pm 0.01$ & $890 \pm 20$ & $-110 \pm 10$ & $0.04 \pm 0.01$ & $30^{+9}_{-8}$ &4700 \Tstrut\Bstrut \\
			& EM4 & $18 \pm 1$ & $ 3.02^{+0.02}_{-0.01}$& $960^{+120}_{-110}$ &$-200^{+150}_{-70}$ & $0.02 \pm 0.01$& $0.8^{+0.5}_{-0.4}$&220 \Tstrut\Bstrut \\
			\hline
			\hline 
			\multirow{5}{*}{2014} & EMA & $33^{+1}_{-2}$ & $3.96^{+0.02}_{-0.01}$  & $2910^{+420}_{-370}$ & $-230^{+30}_{-90}$ & $0.32^{+0.03}_{-0.04}$ & $2.7^{+0.2}_{-0.5}$ & 8600 \Tstrut\Bstrut \\
			& EMB & $51 \pm 1$ & $0.67 \pm 0.01$ & $410 \pm 10$ & $-300 \pm 10$ & $0.02 \pm 0.01$& $510 \pm 260$ & 76100 \Tstrut\Bstrut \\
			& EMC & $29^{+2}_{-1}$ & $1.91 \pm 0.01$ & $920 \pm 20$ & $-230 \pm 10$ & $0.04 \pm 0.01$& $33 \pm 10$& 5740 \Tstrut\Bstrut \\
			& EMD & $37^{+3}_{-2}$ & $3.02^{+0.03}_{-0.02}$  & $1630^{+200}_{-160}$ & $-530^{+150}_{-370}$ & $0.01 \pm 0.01$& $< 1.7$ &  1000 \Tstrut\Bstrut  \\
			\hline
		\end{tabular}
	\end{table*}
	
	\begin{table*}
		\centering
		\caption{Best fit collisionally ionised emission (\texttt{CIE}) component parameters, fitted to the 2000 (top) and 2014 (bottom) RGS and EPIC-PN combined spectra. }
		\label{Table:CIE_Results} 
		\begin{tabular}{c | c c c c | c}
			\hline
			\multirow{2}{*}{Obs.} & EM  & $T_e$ & $v_{out}$  & $v_{t}$  & \multirow{2}{*}{$\Delta C$} \Tstrut\Bstrut \\
			&($10^{69}$ m\textsuperscript{-3}) & (keV) & (km s\textsuperscript{-1}) & (km s\textsuperscript{-1}) & \\
			\hline
			\hline
			2000 (CI1) & $2.72^{+0.08}_{-0.06}$ & $0.57 \pm 0.01$ & $-40^{+30}_{-40}$ & $1220_{-70}^{+50}$ & 1440 \Tstrut\Bstrut \\
			\hline
			2014 (CIA) & $2.19^{+0.06}_{-0.04}$& $0.58 \pm 0.01$ & $-165 \pm 30$ & $1290^{+40}_{-50}$ & 1960 \Tstrut\Bstrut \\
			\hline 
		\end{tabular}
	\end{table*}

	\subsection{2014 best fit}
	\label{Sec:Results_2014}
	
	We fitted first the continuum model (\texttt{POW} and \texttt{REFL}) to the combined spectrum from the 2014 observations, where the initial global fit statistic was C = 125643 (for 1065 d.o.f). All continuum parameters were freed when fitting the emission lines. The change in C-statistic was with respect to the previous component, that was fixed, when introducing the next one, unless stated otherwise. 
	
	For the 2014 spectrum, we also fitted four \texttt{PION} emission components. The first two components, EMA ($\log \xi = 3.96$) and EMB ($\log \xi = 0.67$), improved the fit by $\Delta C_{EMA} = 8600$ and $\Delta C_{EMB} = 76100$, respectively. Fitting these two components together gave a $\Delta C = 2700$. We then fitted a third component (EMC; $\log \xi = 1.91$) which yielded $\Delta C = 5740$. A further $\Delta C = 4670$ was achieved when we freed the parameters in EMA and EMB together with EMC. The fourth, and final, component (EMD; $\log \xi = 3.02$) improved the fit by $\Delta C = 1000$. When we fitted all four \texttt{PION} components together, we achieved a further improvement in the fit of $\Delta C = 540$. 
	
	Similarly to the 2000 spectrum, not all the emission lines were fitted with \texttt{PION} components. We therefore added a \texttt{CIE} component (CIA) to account for the lines produced by a CIE plasma. CIA significantly improved the fit by $\Delta C = 1960$, after we had fixed all of the \texttt{PION} components. Refitting the four \texttt{PION} components with the \texttt{CIE} component further improved the fit by $\Delta C = 760$. 
	
	Next we fitted the abundances, subsequently to fixing all \texttt{PION} and \texttt{CIE} components. The five lower Z elements (C, N, O, Ne, Mg) were fitted in EMB (accounting for the RGS lines) obtaining $\Delta C = 10440$, while the five higher Z elements (Si, S, Ar, Ca, Ni) fitted in EMA improved the fit by $\Delta C = 420$. We then fitted the parameters in each component together, with the abundances free, starting with EMA and EMB,  followed by EMC and EMD, which gave changes in the C-statistic of $\Delta C = 790$ and $\Delta C = 2100$, respectively. We then freed the CIA parameters too and obtained a new best fit with a decrease in C-statistic of $\Delta C = 800$.
	
	
Finally, we accounted for the neutral \ion{Ni}{K$\alpha$} line at 7.5 keV with a Gaussian component (\texttt{GAUS}), fitting the normalisation, energy and broadening velocity. Adding this line to the model improved the best fit by $\Delta C = 440$. Table \ref{Table:Ni_Results} shows the best fit parameter values of the \ion{Ni}{K$\alpha$} line; we were unable to constrain these parameters so we fixed them to their initially fitted values in the subsequent fits and error searches.

	
	All the parameters were fitted together again to obtain a best fit of C = 8510 for 1027 d.o.f. The spectrum and final best fit model, after including more \texttt{GAUS} components (see Sect. \ref{Sec:RGS_PN_Sep}), are shown in Fig. \ref{Fig:RGS_PN_Best_Fit_2014}. The best fit parameter values for the continuum, \texttt{PION}, \texttt{CIE}, and abundances are displayed in Tables \ref{Table:Cont_Results} - \ref{Table:Abund:Results}, respectively. Again, all errors are obtained with all fitted parameters free.

\subsection{Fitting the RGS and PN spectra separately}
\label{Sec:RGS_PN_Sep}
Unfortunately, the fitting statistics for the 2000 and 2014 data are not very good: $\frac{C}{dof} = 4.37$ for 2000, and $\frac{C}{dof} = 8.29$ for 2014. The complex spectrum of NGC 1068, including line blending (\ion{O}{VII} forbidden line with the \ion{N}{VI} RRC between 22 and 23 \AA) and unresolved (for example, \ion{Mg}{XI} triplet at 9 \AA) emission lines, means we are unable to fit all the lines well. For many lines, in particular the He-like triplets, the model underpredicts these features, such that we obtain line ratios that are inconsistent with pure photoionised plasma. This comes down to \texttt{PION} fitting multiple emission lines (with certain $\xi$ and $N_H$ values) simultaneously, rather than individually, making it difficult to achieve a fit that is fully consistent with a photoionised plasma. It is likely that the line-emitting regions in NGC 1068 are more complex than our model can completely match. The geometry and variations in the density and ionisation of the gas, such as clumpy or stratified gas, all affect the overall line spectrum that we get. Nevertheless, this \texttt{PION} model still gives us a more useful insight than an empirical model with Gaussian line fitting. Also, some of the bad fit may be attributed to the instrumental calibration of RGS, which is not perfect. Therefore, we took the best fit models from the simultaneous fits, and folded them to the RGS and PN spectra individually in each epoch, to see if the model is a good fit to individual lines in the separate spectra. 
		
When folding the combined best fit to the PN spectra from each epoch, the statistics were C = 298 for 54 d.o.f ($\frac{C}{dof} = 5.52$) in 2000, and C = 502 for 54 d.o.f ($\frac{C}{dof} = 9.30$) in 2014. This clearly shows a poor fit to the PN spectra only, which count only a total of 94 and 83 bins for 2000 and 2014, respectively. There are residuals between 4 and 6 keV in both epochs, and above 9 keV in 2014. There is also a possible feature at around 5.9 keV (just before the \ion{Fe}{K$\alpha$}), which is also present in the spectrum of \cite{Pounds2006} (their Fig. 3).

In the RGS band, the C-statistic for 2000 was C = 3921 for 994 d.o.f ($\frac{C}{dof}  = 3.94$), whereas for the 2014 data C = 7392 for 994 d.o.f ($\frac{C}{dof} = 7.44$). In the RGS spectra, there were still many residuals where neither \texttt{PION} nor \texttt{CIE} fitted the data well, in particular in the region between 32 and 36 \AA. Therefore, we introduced some Gaussian components (\texttt{GAUS}) to the RGS data only, to improve the fit for these lines. Four \texttt{GAUS} components were fitted to the lines at 35.6, 34.9, 34.3 and 32.2 \AA, while keeping the other model parameters fixed, improving the best fit in the 2000 model by $\Delta C =$ 48, 219, 106, and 66, respectively. In the 2014 data, adding these same four \texttt{GAUS} components, with the other parameters fixed, improved the fit by $\Delta C =$ 105, 466, 169, and 80, respectively. We note that there are no likely identifications with ion species for the lines fitted in the 32 to 36 \AA\ wavelength range. There was also a large residual, in both spectra, at 8.4 \AA, corresponding to the \ion{Mg}{XII} Ly$\alpha$ line, that the \texttt{PION} components only half fitted. We therefore added a \texttt{GAUS} component for this line, improving the fit by $\Delta C = 56$ in 2000 and $\Delta C = 180$ in 2014. Although refitting the two low ionisation \texttt{PION} components and the \texttt{CIE} component to the RGS spectrum (with the \texttt{GAUS} parameters free) did improve the fit in each epoch, there were still many residuals, such as the \ion{O}{VII} triplet between 21.5 and 22.5 \AA. The best fit statistic for the RGS data only was C = 3411 for 973 d.o.f ($\frac{C}{dof} = 3.51$) for 2000, and C = 6356 for 973 d.o.f ($\frac{C}{dof} = 6.53$) for 2014.

As a sanity check, if we model the spectrum locally (between 18 and 23 \AA), we can fit the \ion{O}{VII} triplet and \ion{O}{VIII} Ly$\alpha$ well. But when folding this local model to the global spectrum again (7 - 37 \AA), the emission features outside the 18 - 23 \AA\ range are very poorly fitted, as the parameter values depend on the lines fitted.


\subsection{Final best fit}

Finally, we added these 5 Gaussian components to our models, fitted to the simultaneous RGS and PN spectra, in each epoch. The 2000 and 2014 models were improved by $\Delta C = 520$ and $\Delta C = 1000$, respectively. As these Gaussians did not add to the photoionisation modelling and have no physical meaning, but were introduced to achieve a statistical improvement in the fitted model, we kept the \texttt{GAUS} parameters fixed. We also note that these Gaussian lines do not affect the overall parameter values of the photoionisation modelling.

We then refitted all the parameters together from Tables \ref{Table:Cont_Results}, \ref{Table:PION_Results}, \ref{Table:CIE_Results}, \ref{Table:Abund:Results}, and \ref{Table:Ni_Results}, and obtained final best fit C-statistic values of C = 3790 (for 1037 d.o.f) and C = 7252 (1027 d.o.f), for 2000 and 2014, respectively. Unfortunately, in both epochs the models were still statistically poor: $\frac{C}{dof}  = 3.65$ for 2000 and $\frac{C}{dof} = 7.06$ for 2014. However, due to the complexity of such a rich spectrum, we were unable to improve the fit any further. 
	
In our modelling, this poor fit (in both epochs) may be a result of the assumed geometry of \texttt{PION}, which takes a slab of material and produces forward emission facing us. This simple geometry works very well with type 1 Seyfert AGN because we are observing the AGN and plasma from face on. However, for type 2 AGN, where we see the plasma from side on, the modelling becomes more complex, especially if the outflowing wind is emitting in an ionising cone \citep{Kinkhabwala2002}. Therefore, a simple slab geometry may break down here as the geometry is fundamental for radiation transfer issues, where resonance lines are sensitive to the column density and can vary depending on the LOS. 
This being said, \texttt{PION} is an excellent tool at analysing photoionised plasma self consistently, calculating the photoionisation balance on the fly and computing explicitly the photoionisation properties ($\xi$ and $N_H$) of all the emission lines in the spectrum. However, if the underlying AGN environment is far more complex than assumed by \texttt{PION}, such as an outflowing cone seen in type 2 AGN, then our fitting of the spectra will be simplistic to some extent, explaining why we cannot fit some of the emission lines very well. 

On the other hand, this is a common result for NGC 1068, a Compton-thick AGN, as previous models have not been able to fully explain the Chandra or RGS spectra, showing strong residuals. For example, \cite{Kinkhabwala2002} were unable to model many emission lines between 10 and 16 \AA\ or above 34 \AA\ (see their Fig. 11). \cite{Brinkman2002} were able to fit the Chandra spectrum above 20 \AA\ (in their Fig. 7), but many of the emission lines below 16 \AA\ were not fitted well. Furthermore, \cite{Kraemer2015} were unable to fit the emission lines below 17 \AA\ and above 31 \AA\ (see their Figs. 1 and 4), and \cite{Kallman2014} were unable to model some of the strongest lines such as the \ion{O}{VII} and \ion{Ne}{IX} triplets, again both in the Chandra spectra.

Moreover, none of these modelling attempts accounted for the \ion{Fe}{XVII} lines at 15 and 17 \AA, which we model with a CIE component. This suggests either previous models did not account for photoexcitation, or CIE is a valid explanation. However, \cite{Kinkhabwala2002} did take these \ion{Fe}{XVII} lines into account as due to photoexcitation, although the lines are not fitted in their plot (Fig. 11),  whereas \cite{Brinkman2002} say their model did not include the Fe-L transitions in their model. \cite{Gu2019} tested the contributions to the population of levels by different processes within the \ion{Fe}{XVII} ions, but do not consider photoexcitation. Further tests and comparisons are required in \texttt{SPEX} to fully understand these lines in order to reduce the limitations of photoionisation codes, and to allow \text{PION} to treat photoexcitation properly. Although a \texttt{CIE} component here is able to account for the \ion{Fe}{XVII} lines at 15 and 17 \AA, photoexcitation is just as physically viable, and should therefore be considered as a possibility in NGC 1068. 
	
As a result, given the complexity of this spectrum, the models fitted \citep[by][and here]{Kinkhabwala2002, Brinkman2002, Kallman2014, Kraemer2015} are unable to account for every feature. This means that in all the models the statistical fit is poor and there are many residuals. Only in this paper do we quantify and compare the C-statistic of each component and its contribution to the final model.
	
As a further comparison, the type 1.5 AGN NGC 4151 has a significantly similar RGS spectrum to that of NGC 1068, with many of the same emission features. The modelling of the soft X-ray spectra of NGC 4151 also showed strong under prediction of the continuum and some emission features \citep[see Fig. 3 and Fig. 15 in][respectively]{Schurch2004, Beuchert2017}, implying that it is difficult to obtain an acceptable model statistic, whilst fitting all the emission features in these obscured AGN.

	\subsection{Spectral line features}
	\label{Sec:Line_Features}
	We display the complex spectra from the 2000 and 2014 observations in Figs. \ref{Fig:RGS_PN_Best_Fit_2000} and \ref{Fig:RGS_PN_Best_Fit_2014}, respectively. The best fit model is shown by the red lines, while the other components are represented by different colours (see the legends in each figure). Four \texttt{PION} components in each epoch are required to fit the majority of the emission lines we observe, and we find that components with similar $\xi$ values from each epoch account for the same emission features. For example, EM1 and EMA (green lines) fit the same high energy features, whereas EM2 and EMB (purple lines) fit the same lower energy RGS features, in their respective epochs. This makes it easier when comparing epochs to determine which lines are fitted by each component. This is the same for the \texttt{CIE} components. 
	
	Components EM1 and EMA fit the high energy \ion{Fe}{XXVI} Ly$\alpha$ and Ly$\beta$, and \ion{Fe}{XXV} lines, present in the PN spectrum. EM2 and EMB fit emission lines in the RGS band, such as the \ion{O}{VII} and \ion{N}{VI} triplets, \ion{O}{VII}, \ion{N}{VI}, \ion{C}{VI} and \ion{C}{V} RRCs, and \ion{C}{V} and \ion{N}{VI} H- and He-like species lines. EM3 and EMC fit the \ion{Mg}{XI} triplet lines, \ion{Ne}{IX} RRC, and the \ion{Ne}{IX} He-like species lines, in addition to \ion{O}{VIII} Ly$\alpha$, \ion{Si}{XII}, and \ion{S}{XIII} lines, in both the RGS and PN spectra. EM4 and EMD account for the \ion{Fe}{XVII} and \ion{Fe}{XX} lines in the RGS spectrum, and the \ion{S}{XVI}, \ion{Ar}{XVII}, \ion{Ar}{XVIII}, \ion{Fe}{XXV} lines in the PN data. The \texttt{CIE} component is required to explain the \ion{Fe}{XVII} lines at 15.3 and 17.4 \AA\ (see Fig. \ref{Figure:PION_CIE}).
	
	We find that some emission features are accounted for by many components at the same time. These include \ion{Ne}{X}, \ion{O}{VIII} $\beta$, \ion{N}{VII} and \ion{C}{VI} Ly$\alpha$ lines, \ion{Si}{XIII} and \ion{S}{XV} lines, and the \ion{Ne}{IX} triplet. In addition, some of the emission lines (in the RGS band) are over predicted due to multiple components fitting them. These include the \ion{O}{VIII} and \ion{N}{VI} RRCs, the \ion{Mg}{XI} forbidden line and \ion{Fe}{XVIII} and \ion{Fe}{XIX} lines around 14 \AA.
		
	On the other hand, we do account reasonably well for the features between 10 and 12 \AA\ which were very under predicted by \cite{Kraemer2015}, suggesting that this was because their model had incomplete atomic data. Furthermore, this wavelength region coincides with the failed CCD 7 in RGS1, meaning the effective area has dropped by a factor of two as only data from RGS2 can be analysed between 10 and 15 \AA\ \citep{denHerder2001}. There are still some residuals between 9 and 10 \AA, but this may be a result of multiple lines being blended together.
	
	\subsection{Luminosities}
	We take the ionising luminosity (1 - 1000 Ryd or 13.6 eV - 13.6 keV) for both epochs to be $L_{ion} = 1.54 \pm 0.04 \times 10^{37}$ W, calculated from the assumed ionising SED\footnote{Similar ionising luminosity compared to NGC 7469, the only difference comes from the \texttt{REFL} component that we fit here.}; we use this value for further calculations (Sect. \ref{Sec:Distances}). In addition, we calculate the observed 2 - 10 keV luminosity (using \texttt{POW} and \texttt{REFL} parameters in Table \ref{Table:Cont_Results}) from our best fit model, where we obtain $L^{2000}_{2-10} = 1.07^{+0.06}_{-0.04} \times 10^{34}$ W and $L^{2014}_{2-10} = 1.04^{+0.04}_{-0.03} \times 10^{34}$ W, for the 2000 and 2014 observations, respectively. The intrinsic 2 - 10 keV luminosity unveiled during a temporary decrease in absorption was $L_{obs,\ 2 - 10} = 7^{+7}_{-4} \times 10^{36}$ W \citep{Marinucci2016}. By taking the ratio between our observed luminosities and the intrinsic luminosity, we find that roughly 0.15 per cent of the intrinsic X-ray source is reflected and scattered into our LOS.	
	
	\section{Comparing epochs}
	\label{Sec:Comp_Epochs}
After fitting the simultaneous RGS and EPIC-PN spectra of NGC 1068 from 2000 and 2014, we find almost no difference in the emission features and component properties between each epoch. Here we discuss our overall findings and results from our photoionisation modelling.

	\subsection{Observed continuum}
	
	Starting with the observed continuum, both the power-law and reflection components differ between epochs (see Table \ref{Table:Cont_Results}). The photon index is flatter in 2014, while there is a decrease in normalisation for both \texttt{POW} and \texttt{REFL} components compared to 2000; they are not consistent within the uncertainties. We are able to constrain the line width ($\sigma_v$) of the neutral \ion{Fe}{K$\alpha$} line (measured by coupling a \texttt{VGAU} component to \texttt{REFL}) in 2000, but obtain an upper limit for 2014.
	
	
	\subsection{Photoionised plasma components}
	
For the four \texttt{PION} components, we find very little difference in the ionisation parameters between the two observations (not significant given the very little change in spectra). This suggests that the components with the same ionisation parameter in each epoch are the same plasma region. On the other hand, we do find the total equivalent column density increases from $N_H^{2000} = 131^{+5}_{-6} \times 10^{25}$ m\textsuperscript{-2} in 2000 to $N_H^{2014} = 150^{+7}_{-6}  \times 10^{25}$ m\textsuperscript{-2} in 2014. This change of $\Delta N_H = 19 \times 10^{25}$ m\textsuperscript{-2} is fairly significant, with an average of $\Delta C \sim 3900$, when substituting the $N_H$ values from one epoch into the model of the other. Below we describe some possible reasons for this increase in total equivalent column density, by comparing each component of similar ionisation state.

A degeneracy was found between $N_H$ and $C_{cov}$ by \cite{DiGesu2017} in their analysis of the Seyfert 1 AGN 1H 0419-577, using \texttt{PION}. This may explain the decrease in column density ($\Delta N_H = - 17 \times 10^{25}$ m\textsuperscript{-2}) between EM1 and EMA, where there is an increase in covering fraction ($\Delta C_{cov} = 0.13$). For EM4 and EMD, the change in column density ($\Delta N_H = + 19 \times 10^{25}$ m\textsuperscript{-2}) between the two epochs may actually be due to a degeneracy between $N_H$ and $v_{turb}$. In this case there is a small change in covering fraction ($\Delta C_{cov} = 0.01$), but the turbulent velocity has increased by $\Delta v_{turb} = 670$ km s\textsuperscript{-1} (the uncertainties of each parameters are smaller than this difference). This may be due to the large turbulent velocities being driven by line ratios, rather than line widths, causing changes in $N_H$ between components and epochs. In addition, the curve-of-growth illustrates how both $N_H$ and $v_{turb}$ (line broadness) can affect the intensity of the lines. For EM2 and EMB, and EM3 and EMC, the changes in column density are $\Delta N_H = + 10 \times 10^{25}$ m\textsuperscript{-2} and $+ 7 \times 10^{25}$ m\textsuperscript{-2}, respectively. The change in $C_{cov}$ is negligible, but the $v_{turb}$ values increase by 40 and 30 km \textsuperscript{-1}, respectively for EM2/B and EM3/C; this difference is larger than the uncertainties on the parameters for EM2/B. The column density changes for each individual component between epochs overall cause a total increase in equivalent column density of $\Delta N_H = 19 \times 10^{25}$ m\textsuperscript{-2}. The varying $N_H$ values between the two epochs are unlikely to be real, however, as there is very little spectral change between the two epochs (Fig. \ref{Fig:Spec_Comparison}). Therefore, the line column densities cannot vary as greatly as suggested here by the fit, and the most probable explanation for the increase in total $N_H$ from 2000 and 2014 is model degeneracy with $v_{turb}$ and $C_{cov}$.
	
	The $C_{cov}$ values for EM1 and EMA are significantly larger than the other three components, suggesting that this plasma region is closer to the central black hole than the other components. This is also consistent with the high ionisation parameter. However, these values are significantly smaller compared to the covering factor values of the two RGS components found by \cite{Kraemer2015}. Furthermore, the broadening velocities ($v_{turb}$) in Table \ref{Table:PION_Results} of both EM1 and EMA are consistent with the broadening velocity of the Balmer lines measured at $\sigma_{Balmer} \sim 3200$ km s\textsuperscript{-1} \citep{Antonucci1985}. These measurements indicate that the highly ionised components are close to the central SMBH, possibly consistent with the BLR, although we cannot observe the BLR directly due to obscuration from the torus. So for now, we assume all plasma regions are part of the NLR. On the other hand, if the X-ray BLR emission is largely reflected \citep[like the optical BLR emission;][]{Antonucci1985}, then the BLR would be a possible origin for the highly ionised emission lines. 
	
	In addition, the emission measures (EM) of each \texttt{PION} component are consistent between the two epochs. The definition of the EM is given by $\int n_e n_H dV$, but as the number densities of electrons ($n_e$) and ions ($n_H$), respectively, are unknown, we use the following equation \citep[e.g.][]{Mao3783, Grafton-Waters2020}
	\begin{equation}
	EM = \frac{n_e}{n_H}\frac{4\pi N_H C_{cov} L_{ion}}{\xi},
	\label{Eq: E.M.}
	\end{equation}
	where $n_e / n_H \sim 1.2$ for fully ionised plasma and $L_{ion}$ is the ionising SED luminosity $L_{ion} = 1.54 \times 10^{37}$ W. For the \texttt{PION} components, if we assume a constant $L_{ion}$ and $C_{cov}$, then EM depends on $N_H$ and in particular $\xi$. If $\xi$ is small, then the number density in the plasma is larger (for a fixed distance), which means EM (from the definition) increases, and thus explains why the lower ionised plasmas have the highest EM values. The EM, $\xi$ and $N_H$ values for these four components in each epoch (Table \ref{Table:PION_Results}) are consistent with the model components from \cite{Kallman2014}.

	\subsection{Collisionally ionised plasma}
	\label{Sec:CIE_Plasma}
	The electron temperatures ($T_e$) and line width ($v_t$) for each \texttt{CIE} component have not changed (within errors) between the two epochs, but the EM and $v_{out}$ values are not consistent within the errors (see Table \ref{Table:CIE_Results}). However, much like with the \texttt{PION} components, it is the outflow velocities that differ the most. In 2000, the outflow velocity is negligible with $v_{out} = -40^{+30}_{-40}$ km s\textsuperscript{-1}, whereas in 2014 the CIE plasma appears to be travelling at $-165 \pm 30$ km s\textsuperscript{-1}. 
	
	Although the \texttt{CIE} components in each epoch account for the \ion{Fe}{XVII} lines that \texttt{PION} cannot fit\footnote{\texttt{PION} is able to fit the low Z ions (like oxygen, carbon, nitrogen) but not high Z ions such as iron; hence why we need a \texttt{CIE} component.}, this is not an overall convincing conclusion. One argument against CIE is that the RGS spectra show strong, narrow RRC features, implying the X-ray emission is only from PIE plasma, making CIE unlikely in NGC 1068. However, RRC emission could come from outflowing PIE plasma closer to black hole, while the CIE emission could originate from further out, such as the secondary region or star burst region. In terms of the Fe-L lines themseleves, the \ion{Fe}{XVII} line at 17 \AA\ is populated by recombination rather than direct excitation, so PIE plasma should dominate this line. As for the \ion{Fe}{XVII} resonance line at 15 \AA, the 3d shell is highly populated if the plasma density is high (in a similar way the resonance line in He-like triplets is more dominant over the forbidden line if the density is large), suggesting that CIE plasma can account for this. Figures 1 and 2 from \cite{Liedahl1990} show that the \ion{Fe}{XVII} at 17 \AA\ should be accounted for by PIE plasma, whereas the 15 \AA\ line is emitted in CIE plasma, reiterating the arguments above. However, photoexcitation was not considered by \cite{Liedahl1990}, and would explain the 15 \AA\ resonance line \citep{Sako2000, Kinkhabwala2002}; in this case the explanation for the lack of PIE emission at 17 \AA\ is, however, unknown. Therefore, the under-prediction of these lines by \texttt{PION} suggests we are not getting the full picture, and although CIE plasma can produce these \ion{Fe}{XVII} lines (Fig. \ref{Figure:PION_CIE}), photoexcitation is a more natural process in the outflowing wind \citep{Kinkhabwala2002}. Although we fit a \texttt{CIE} component to each epoch to account for the \ion{Fe}{XVII} lines, photoexcitation in PIE plasma cannot be dismissed.

	\subsection{Abundances}
	We fitted abundances for ten elements that are present in both RGS and PN spectra of NGC 1068 \citep[from][respectively]{Kinkhabwala2002, Pounds2006}. Between the two epochs, there is very little difference in the ratios with respect to Fe as shown in Table \ref{Table:Abund:Results}.
	\cite{Kinkhabwala2002} and \cite{Brinkman2002} found the N abundance to be 2-3 times Solar, however we do not find any evidence for this (see Table \ref{Table:Abund:Results}).

\subsection{The \ion{Ni}{K$\alpha$} line}
Table \ref{Table:Ni_Results} shows the best fit values for the \ion{Ni}{K$\alpha$}, modelled with a \texttt{GAUS} component. However, for the 2014 line, we are unable to constrain the parameters of this \texttt{GAUS}, so we fix them to the values initially fitted. As a comparison, \cite{RB2020} constrained the width of the \ion{Fe}{K$\alpha$} and \ion{Ni}{K$\alpha$} lines to be of the order of 300 km s\textsuperscript{-1}, lower than what we quote here from EPIC-PN. This could be due to the HETGS instrument having a higher spectral resolution than EPIC-PN.

		\begin{table*}
	\centering 
	\caption{Best fit abundances with respect to iron. \textit{Top:} abundances fitted to the 2000 observation. EM2 accounts for the elements in the RGS energy range while EM1 fits those in the PN range. \textit{Bottom:} abundances fitted to EMB for the RGS spectrum and EMA for PN energies in the 2014 spectrum. The abundances of the other components are coupled to these values, such that all the lines are fitted by the ions in each component, as we assume that the chemical enrichment is the same throughout the nucleus of NGC 1068.}
	\label{Table:Abund:Results}
	\begin{tabular}{c | c | c c c c c | c  }
		\hline
		\hline
\multirow{5}{*}{2000} & \multirow{2}{*}{EM2 (RGS)}& C & N & O & Ne & Mg & $\Delta C$ \Tstrut\Bstrut \\
& & $0.31 \pm 0.01$& $0.67 \pm 0.01$ & $0.15 \pm 0.01$ & $0.32 \pm 0.01$ & $0.38^{+0.01}_{-0.02}$ & 5200 \Tstrut\Bstrut \\
\cline{2-8}
& \multirow{2}{*}{EM1 (PN)} & Si & S & Ar & Ca & Ni & $\Delta C$ \Tstrut\Bstrut \\
& &$0.47 \pm 0.01$ & $0.58 \pm 0.02$ & $ 0.83^{+0.06}_{-0.05}$ & $< 0.04$  & $< 0.01$& 70\Tstrut\Bstrut\\
\hline \hline
\multirow{5}{*}{2014}&  \multirow{2}{*}{EMB (RGS)} & C & N & O & Ne & Mg & $\Delta C$  \Tstrut\Bstrut \\
& & $0.27 \pm 0.01$ & $0.59^{+0.05}_{-0.06}$ & $0.13 \pm 0.01$ & $0.30 \pm 0.01$ & $0.36 \pm 0.01$ & 10600 \Tstrut\Bstrut \\
\cline{2-8}
&\multirow{2}{*}{EMA (PN)} & Si & S & Ar & Ca & Ni & $\Delta C$ \Tstrut\Bstrut \\
& &$0.42 \pm 0.01$ & $0.63^{+0.02}_{-0.01}$ & $0.67^{+0.04}_{-0.03}$& $< 0.01$& $< 0.01$ & 350 \Tstrut\Bstrut \\
\hline
	\end{tabular}
\end{table*}

\begin{table}
	\centering
	\caption{Best fit \ion{Ni}{K$\alpha$} parameter values for 2000 (left) and 2014 (right) spectra.}
	\label{Table:Ni_Results}
	\begin{tabular}{c | c c}
		\hline
		Parameter & 2000 & 2014 \\
		\hline
		N ($10^{47}$ ph s\textsuperscript{-1}) & $2450^{+310}_{-550}$ & $3600$ (f) \Tstrut\Bstrut\\
		E (keV) & $7.50^{+0.01}_{-0.22}$ & $7.47$ (f) \Tstrut\Bstrut\\
		$\sigma_v$ (km s\textsuperscript{-1}) & $4040^{+1070}_{-2170}$ & $7000$ (f) \Tstrut\Bstrut\\
		\hline
		$\Delta C$ & 130 & 440 \\
		\hline
	\end{tabular}
\tablefoot{The 2014 parameter values could not be constrained, so we fixed (f) them to their initial fitted values.}
\end{table}

	\section{Discussion}
	\label{Sec:Discussion}
	In this section we discuss the results from our spectral modelling of NGC 1068 in 2000 and 2014. Here we use our data to infer and determine properties of the emitting plasma that surrounds the nucleus in this AGN. In particular, we investigate the distances of each photoionised component from the SMBH, and determine the thermal stability of the plasma regions for the two epochs, before studying the validity of the CIE emission. 

\subsection{Photoionised plasma distances}
\label{Sec:Distances}

After determining our best fit photoionisation models for the NGC 1068 spectra in both epochs, we now want to establish the locations of the emitting plasma regions, with respect to the central SMBH. Unfortunately, as seen in Fig. \ref{Fig:Spec_Comparison} and Table \ref{Table:PION_Results}, there is no variability between the two epochs, either in terms of the spectral features or the ionisation state of the plasma components, suggesting that the emitting plasma has not changed over the 14 year period between observations.
This means that we are unable to obtain distance measurements from variability arguments where a change in the ionisation parameter is due to a change in the SED shape \citep[see e.g.][for NGC 7469]{Mehdipour2018}. In addition, as we are unable to directly observe the ionising X-ray source, attaining an intrinsic SED is practically impossible. Therefore, we have to investigate alternative ways to calculate the distances of the plasma regions, discussing the physicality, as well as the advantages and negatives, of each method. These measurements depend on the geometry and our LOS view of NGC 1068, which adds complexity in comparison to Seyfert 1 AGN.

The uncertainties on the distance estimates here are obtained from the parameters and errors of our best fit modelling using \texttt{PION} (Table \ref{Table:PION_Results}). The hydrogen number densities for each component, and their respective errors, are shown in Table \ref{Table:Distance_Comp}; we multiply these values by 1.2 to get the electron number density that we use in our calculations. Ideally, the plasma density should be constrained through the recombination timescale using the variability between observations, but as this is not the case here, we have used the calculated densities from \texttt{PION}.
	
	\subsubsection{Ionisation parameter distances}
	\label{Subsec:Ion_Dist}
	We start by estimating the locations of these emission components using the definition of the ionisation parameter ($\xi$)	
	\begin{equation}
	\xi = \frac{L_{ion}}{n_e R^2},
	\label{Eq:ION}
	\end{equation}
	where $n_e$ is the electron number density, $L_{ion}$ is the ionising luminosity (measured between 1 - 1000 Ryd) and $R$ is the distance of the plasma from the black hole. From the fits with \texttt{PION}, we are able to derive the hydrogen number densities ($n_H$) \footnote{\texttt{PION} calculates the hydrogen density by fitting the line ratios of all density-sensitive lines in the spectrum (taking into account processes like resonant scattering).} for each emission component (displayed in Table \ref{Table:Distance_Comp}). Obtaining the densities allows us to estimate the distances of each component; alternatively, \cite{Peretz2019} used the R ratios and plots from \cite{Porquet2000} to obtain the number density for the BLR in NGC 4051. 
	
	Unfortunately, in both these methods (and the method below in Sect. \ref{Subsec:Size_Dist}), the number density values of EM1 and EMA cannot be accurately determined. Therefore the distance estimates cannot be obtained for these two components. For EM4 and EMD, we are unable to constrain the values and instead have density upper limit values. For the other two components in each epoch, the number densities are constrained.
	
	Using the ionising luminosity of $L_{ion} = 1.54 \times 10^{37}$ W (for both epochs) and the ionisation parameters from Table \ref{Table:PION_Results} we can estimate the distances for each component using Eq. \ref{Eq:ION}. Alternatively, we can obtain a similar answer using the EM from Eq. \ref{Eq: E.M.} and substituting it into the following equation \citep{Mao5548}
	\begin{equation}
	R^2 = \frac{EM}{4 \pi n_e C_{cov} N_H },
	\label{Eq:Dist_EM}
	\end{equation}
	where $N_H$ and $C_{cov}$ are values from Table \ref{Table:PION_Results}. However, this equation cancels down to Eq. \ref{Eq:ION} when we substitute EM from Eq. \ref{Eq: E.M.}, but with a factor of 1.2 difference between $n_e$ and $n_H$.
	
These methods provide a first distance indication, but work best when the intrinsic SED is known and when there is large variability between observations. The resulting distances ($R_{\xi}$ and $R_{EM}$, respectively) are listed in Table \ref{Table:Distance_Comp}.

	\subsubsection{Using the size of the component}
	\label{Subsec:Size_Dist}
	The second method considers the size of each component estimated from the ratio between the column density ($N_H$) and electron number density ($n_e = 1.2 n_H$ for fully ionised plasma). We then use the relation between the covering fraction ($C_{cov}$) and the solid angle ($\Omega$) to find the area $A$ of each component as seen from the black hole: 
	\begin{equation}
	C_{cov} = \frac{\Omega}{4\pi} = \frac{A}{4\pi R^2}
	\label{Eq: C_cov_1}
	\end{equation}
	where \textit{R} is the distance from the black hole. Here, we assume that each emitting component is spherical and the black hole sees only half of the sphere with area $A = 2\pi r^2$, where r is the radius of the component. However, the radius of each sphere, which is its characteristic thickness, is $\Delta r \approx \frac{N_H}{n_e}$ \citep[identical to Eq. 6 in][]{Peretz2019}, and therefore $A = \pi \Delta r^2$.
	
	Substituting \textit{A} into Eq. \ref{Eq: C_cov_1} and rearranging for R gives the distance from the black hole, given the size of each component
	\begin{equation}
	R^2 \approx \frac{\Delta r^2}{4 C_{cov}}.
	\label{Eq: C_cov_2}
	\end{equation}
	With this method it is fairly easy to obtain distance measurements based on the geometry between the black hole and plasma region. In addition, Eq. \ref{Eq: C_cov_2} does not depend directly on the SED, although $N_H$ and $n_e$ are measured with \texttt{PION} which does require the correct SED input. However, the $N_H/n_e$ ratio gives an approximation for the thickness of a slab of emitting plasma, whereas we assume each component is spherical. Our assumption that each component is spherical is a crude one, and an over simplification of the plasma geometry that does not relate easily to the thickness. Maybe a more appropriate geometry for the plasma is a cube, with dimensions equal to the thickness $N_H / n_e$. In addition, the column density (and number density) are measured with respect to our LOS, both of which are dependent on the viewing angle. This model relies heavily on the size and geometry of each plasma component, which is difficult to calculate with certainty, especially when viewed from side on. Distance estimates obtained with this method ($R_{\Delta r}$) are shown in Table \ref{Table:Distance_Comp}, except for EM1 and EMA due to inaccurate $n_e$ values. 
	
	\subsubsection{Escape velocity distances}
	\label{Subsec:Vel_Dist}
	Next we consider the escape velocity of each component, and assume the outflow velocities from Table \ref{Table:PION_Results} are large enough to allow the plasma components to escape the gravitational potential of the black hole. Therefore, setting $v_{out} = v_{esc}$ we can obtain distances from 
	
	\begin{equation}
	R = \frac{2 G M_{BH}}{v^2_{out}},
	\label{Eq:vel_dist}
	\end{equation}
	where $G$ is the gravitational constant and $M_{BH} = 1.6 \times 10^7 M_{\odot}$ is the mass of the black hole \citep{Panessa2006}. 
	
	This method is independent on the SED. However, the outflow velocity, $v_{out}$, is along our LOS and therefore depends on the viewing angle with respect to a face on view (type 1) AGN. Although the outflowing wind makes a cone-like shape, the $v_{out}$ we measure is not the net outflow velocity and depends strongly on wavelength accuracy of the RGS spectrum \citep[$\delta \lambda = 8$ m\AA;][]{denHerder2001}. In addition, while Eq. \ref{Eq:vel_dist} was used by \cite{Blustin2005} to obtain minimum distance limits on the WA components in AGN, the $R_{v_{out}}$ values we derive are not the lowest distance measurements for each component in Table \ref{Table:Distance_Comp}.

	\subsubsection{Estimated volume filling factor distances}
	\label{Subsec:Vol_Dist}
	In NGC 7469, both the WA \citep{Blustin2007} and emission line region \citep{Grafton-Waters2020} distances were estimated using the volume filling factor $f_v$. The $f_v$ values were calculated for the WA components, but the values for the emission line regions had to be fixed at some arbitrary values for the NLR and BLR \citep[see][and references within]{Grafton-Waters2020}. Ideally, we would want to calculate the $f_v$ values for each individual emission component within NGC 1068. 
	
The volume filling factor equation was derived from the mass outflow rate by \cite{Blustin2005}. They calculated the distances of the WA in many type 1 AGN by assuming that the momentum of the outflowing wind is related to the momentum of radiation being absorbed ($P_{abs}$) plus the momentum of the scattered radiation ($P_{scat}$), which depends on the size of each plasma region \citep{Blustin2005}. The equation is given by
	\begin{equation}
	f_v = \frac{(\dot{P}_{abs} + \dot{P}_{scat})\xi}{1.23 m_p L_{ion} v^{2}_{out} 4 \pi C_{cov}},
	\label{Eq:Dist_Cv}
	\end{equation}
	where $m_p$ is the proton mass, and the remaining parameters are calculated in the best fit models for each epoch.  
	
	This equation works well for the WA components in type 1 AGN, such as NGC 7469, because our LOS is approximately aligned with the wind's direction of motion, and the radiation, which drives the wind (giving it momentum), can be measured directly. However, in type 2 AGN, the direction of the plasma motion is almost perpendicular to our LOS, and we are unable to measure the intrinsic luminosity in order to determine, for example, the two momenta in Eq. \ref{Eq:Dist_Cv}. This, therefore, means we are unable to obtain accurate $f_v$ values for the components within NGC 1068.
	
	Instead we set values for the volume filling factor. The volume filling factor has been quoted to be between 0.001 and 0.01 \citep[e.g.][]{Osterbrock1991, Snedden1999} for the BLR, but there is no information regarding the NLR. In NGC 7469, \cite{Grafton-Waters2020} set $f_v = 0.1$ for the two NLR components, such that the emission line regions were further from the black hole than the WA components, and $f_v = 0.001$ for the BLR component. In NGC 1068, we do not measure any absorption due to our LOS view \citep[e.g. the cone model in][]{Kinkhabwala2002}, but we initially set $f_v = 0.1$ for all \texttt{PION} components, although \cite{Kallman2014} suggested a volume filling factor of 0.01.

	Components EM1 in 2000 and EMA in 2014 have large $C_{cov}$ values relative to the other components, suggesting they are more extended as seen by the SMBH, and are therefore closer (see Eq. \ref{Eq: C_cov_2}). The broadening velocities ($v_{turb}$) in Table \ref{Table:PION_Results} of these components are consistent with those of the Balmer lines in NGC 1068 \citep[$\sigma_{Balmer} \sim 3200$ km s\textsuperscript{-1};][]{Antonucci1985}, and with the line broadening velocities for BLRs found in type 1 AGN \citep[e.g.][]{Kollatschny2013}. Furthermore, the ionisation parameters of EM1 and EMA are very large, and these components only fit the high energy lines in the PN spectra for both epochs, suggesting that these lines are produced from highly ionised plasma, closer to the central black hole than the other components.
	
	The parameter values point us to the conclusion that EM1 and EMA could be part of the BLR. However, we cannot see the emission from the BLR because the nucleus in NGC 1068 is obscured by the torus. In optical observations, the BLR is reflected off the far side of the dusty torus, or scattered off free electrons, where the light we observe is polarised, relative to the unreflected emission \citep{Antonucci1985}. The amount of polarisation of the BLR emission is consistent to that of the underlying continuum (in our case modelled with \texttt{REFL}), whereas the NLR shows negligible polarisation \citep{Antonucci1985}. This could also be true for the X-ray BLR whereby the emission lines are reflected into our LOS. However, there is no significant evidence in our modelling that suggests the emission is being reflected into our LOS, so we cannot say conclusively that we are seeing emission from the BLR. Therefore, we assume both EM1 and EMA are observed directly and are part of the NLR.
	
	
	There are two methods that can be used to obtain distance estimates using the volume filling factor. Both methods start with $N_H = n_e f_v \Delta r$ and $\xi = L_{ion}/n_e R^2$, where $n_e$ is the electron number density, $\Delta r$ is the thickness of the plasma region and R is the plasma distance from black hole; $L_{ion}$, $\xi$, and $N_H$ are from our modelling. The first derivation, from \cite{Blustin2005} \citep[and used on NGC 7469 by][]{Blustin2007}, assumed a thin layer (of thickness $\Delta r$) for each plasma component containing most of the mass, with an ionisation $\xi$. This method led to a maximum distance limit because \cite{Blustin2005} assumed $\Delta r \leq R$. Alternatively, \cite{Grafton-Waters2020} \citep[see also][]{Behar2003, Whewell2015} derived the distance estimate by integrating $n_e$ (substituting for $n_e = \frac{L_{ion}}{\xi R^2}$; Eq. \ref{Eq:ION}) over the thickness of the plasma region such that $N_H = \int \frac{L_{ion} f_v}{\xi R^2} dR$, where they assumed the plasma outer distance to be further from the black hole than the inner distance. Therefore, this second method measured a minimum distance value. The values calculated with these two methods are the same, it just depends on the method used as to whether an upper distance limit or a minimum distance is achieved. Therefore, for ease, we shall assume this equation gives the locations of the emission components,
	
	\begin{equation}
	R = \frac{L_{ion} f_v}{\xi N_H}.
	\label{Eq:vol_fill_dist}
	\end{equation}
	
	For NGC 1068, $f_v$ cannot be measured for the plasma regions, therefore we set $f_v$ to the same arbitrary values for each component in 2000 and 2014. Consequently, there are very large uncertainties associated with $f_v$, which means the errors on the distances obtained from this method are significantly greater than what is derived from our fitting and what is quoted in Table \ref{Table:Distance_Comp}. The evident discrepancies between the result of this distance method and the others are likely to come from these large $f_v$ uncertainties, which may be up to orders of magnitude in difference. Furthermore, Eq. \ref{Eq:vol_fill_dist} requires two \texttt{PION} parameters ($N_H$ and $\xi$) as well as the ionising luminosity, none of which are secure because we do not know the true shape of the SED, adding to the uncertainties on the distance values. Distance estimates obtained from this method ($R_{f_v}$) are shown in Table \ref{Table:Distance_Comp}.
	
	\subsubsection{Comparing distance estimates}
	\label{Sec:Comp_Dist}
\begin{figure*}
		\centering
		\begin{subfigure}{0.8\linewidth}
			\includegraphics[width=1\linewidth]{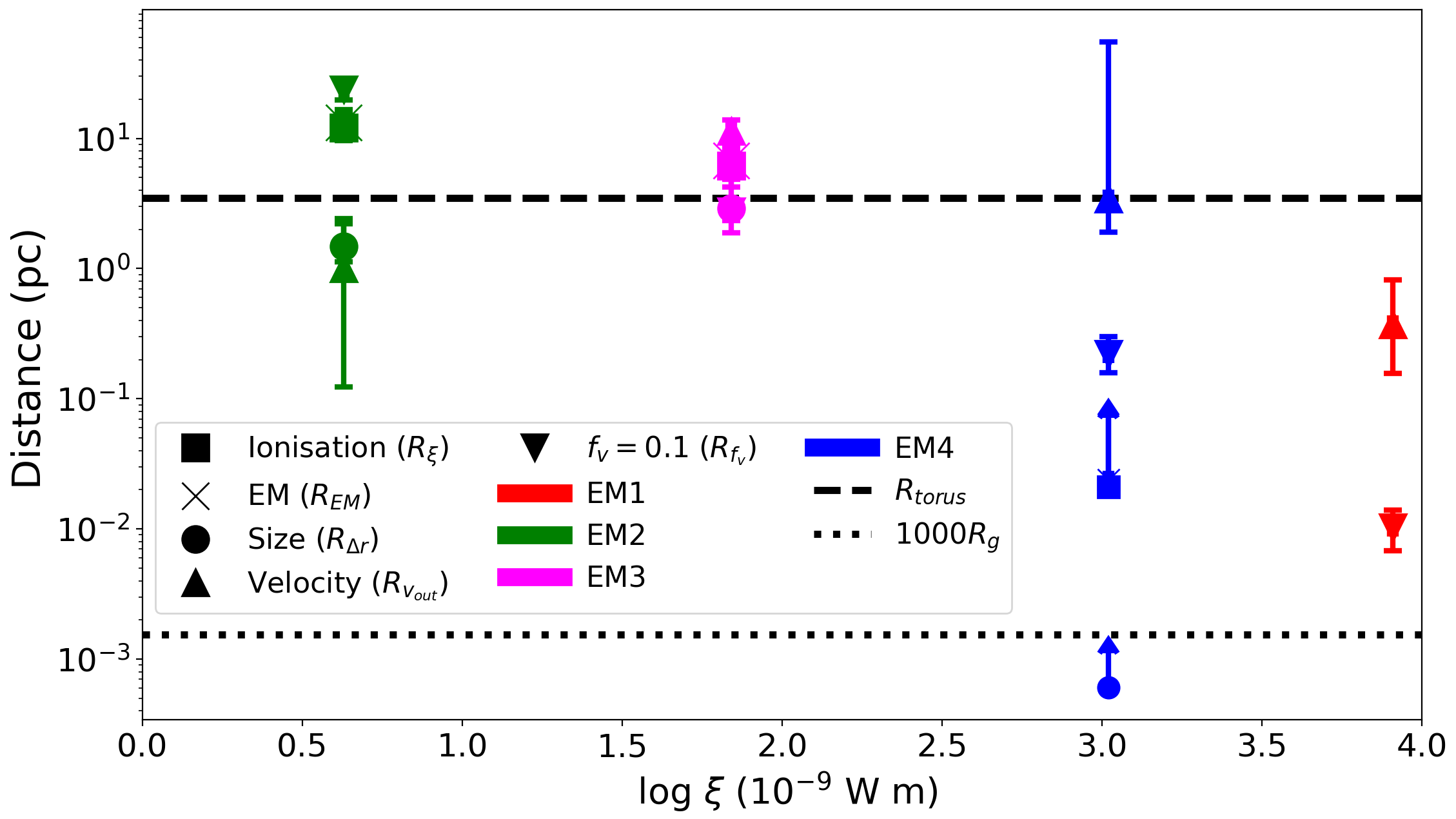} 
		\end{subfigure}
		\begin{subfigure}{0.8\linewidth}
			\includegraphics[width=1\linewidth]{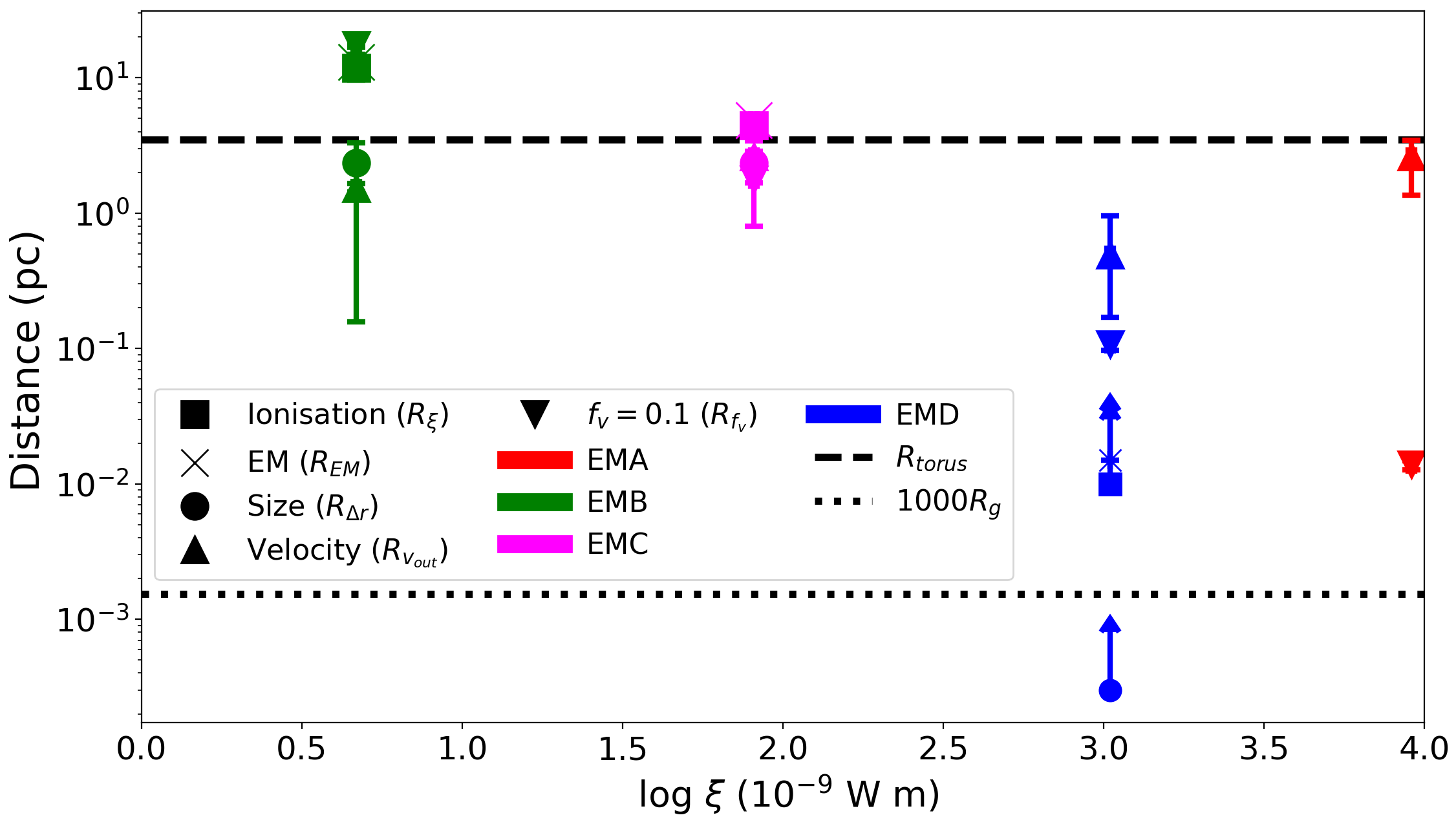}
		\end{subfigure}
		\caption{Distances of each component are plotted for the different methods in Sect. \ref{Sec:Distances}, at the respective ionisation parameters for 2000 (top) and 2014 (bottom). The data point shapes are shown in the legend, as are the colours for each component. The dashed line in each plot represents the location of the torus, at $r_{torus} = 3.5$ pc \citep{Garcia-Burillo2016}, while the dotted line is at 1000 gravitational radii ($R_g = \frac{2GM_{BH}}{c^2}$) from the SMBH, corresponding to a typical outer disk radius \citep{Netzer2013}.}
		\label{Fig:Comp_Distances}
	\end{figure*}

Figure \ref{Fig:Comp_Distances} illustrates the estimated distances from the central SMBH, as a function of ionisation parameter of each component, while Table \ref{Table:Distance_Comp} displays them quantitatively.  
	The distances vary because the parameters and assumptions used in each method are different, in addition to using a non-intrinsic SED in our photoionisation modelling. Although we are unable to constrain the distances shown in Fig. \ref{Fig:Comp_Distances}, and a range of estimated values for each component is found, these findings are very useful, with some components being within the same distance ranges as each other, suggesting they are part of the same emitting region \cite[as found by][]{Kallman2014}. Overall, the distances for each component are consistent with the expectations based on the values in Table \ref{Table:PION_Results}.
	
Fig. \ref{Fig:Comp_Distances} shows that the most ionised components, EM1 and EMA ($\log \xi \sim 4$), and EM4 and EMD ($\log \xi \sim 3$) are closest to the black hole, compared to the low ionised components. For EM1 and EMA, due to unconstrained number densities, we only have two distance estimates. The larger column density and outflow velocity of EM1 puts it closer to the black hole compared to EMA. In both epochs, the distances are within the torus radius, suggesting they could be part of the BLR. EM4 and EMD have the largest range in distance estimates, mostly due to the upper limits of the number density and therefore a lower distance limit. The method using the plasma size ($R_{\Delta r}$) places EM4 and EMD close to the black hole, with the number density upper limit consistent with the BLR ($n_{BLR} \sim 10^{15}$ m\textsuperscript{-3}). However, this location is probably unphysical as the high density limit is a result of EM4 and EMD not fitting many emission lines (Figs. \ref{Fig:RGS_PN_Best_Fit_2000} and \ref{Fig:RGS_PN_Best_Fit_2014}) and statistically improving the C-statistic the least. The consistencies in distances between high ionisation components, EM1 and EM4, and EMA and EMD, suggest that they are at similar locations; this is also shown by the stability plots in Fig. \ref{Fig:Thermal_Plots} (see Sect. \ref{Sec:Thermal_Stability}).

	For the low ionised components in each epoch, EM2 and EMB are furthest from the black hole, and have a large range of distance estimates, placing them either side of the torus, depending on the method used. For example, the large outflow velocity (compared to the other components) places them close to the black hole, while the low ionisation state indicates they are further away. EM3 and EMC, on the other hand, have the most constrained distance estimates, placing the plasma region consistent with the torus distance \citep[$R_{torus} = 3.5$ pc;][]{Garcia-Burillo2016}. These components have the lowest outflow velocities in each epoch, along with a low column density for a moderate ionisation state of $\log \xi \sim 2$. In addition, we can constrain the number density of EM3 and EMC.
	
	None of the components are found at the location of the secondary region of X-ray emission at 290 pc \citep[which extends to about 400 - 500 pc from the central black hole;][]{Brinkman2002, Ogle2003, Garcia-Burillo2017}. In addition, the secondary region has a column density three times smaller compared to the primary region \citep{Brinkman2002, Ogle2003}, but none of the equivalent column densities of the four components (Table \ref{Table:PION_Results}) are a third of the size of any other, again suggesting none of these components, in either epoch, are situated outside the circumnuclear disk. 
	
	\begin{sidewaystable}[]
		\centering
		\caption{Distance values calculated from the various methods in Sect. \ref{Sec:Distances}. We also note the hydrogen number density ($n_H$) and component size ($\Delta r$) required to calculate some distances.}
		\label{Table:Distance_Comp}
		\begin{tabular}{c | c|c c c|  c  c | c c}
					\hline 
		Obs. & Comp. & $n_H$ (m\textsuperscript{-3}) & $R_{\xi}$ (pc) &  $R_{EM}$ (pc) & $\Delta r$ (pc) & $R_{\Delta r}$ (pc) &  $R_{v_{out}}$ (pc)  &  $R_{f_v}$ (pc) \tablefootmark{$\dagger$} \Tstrut\Bstrut \\
		\hline
		\hline 
		\multirow{5}{*}{2000}&  EM1 & -\tablefootmark{*} & - & -  & - & -  & $0.37^{+0.45}_{-0.22}$ & $0.010^{+0.004}_{-0.003}$ \Tstrut\Bstrut \\
		& EM2 & $2.17^{+1.80}_{-0.76} \times 10^{10}$ &  $12.06^{+3.44}_{-2.46}$&  $13.22^{+3.50}_{-1.58}$ & $0.51^{+0.27}_{-0.12}$  & $1.43^{+0.94}_{-0.35}$ & $1.01^{+1.20}_{-0.89}$ &  $23.75^{+2.28}_{-4.06}$ \Tstrut\Bstrut \\
		& EM3 & $5.11^{+4.81}_{-2.49} \times 10^9$ &  $6.17^{+2.03}_{-1.92}$ & $6.76^{+2.22}_{-1.54}$ & $1.16^{+0.65}_{-0.59}$  & $2.91^{+1.93}_{-1.02}$ &  $11.43^{+2.40}_{-1.83}$  & $2.73^{+0.03}_{-0.04}$ \Tstrut\Bstrut \\
		& EM4 & $< 2.83 \times 10^{13}$ & $> 0.021$ &  $> 0.024$ & $> 0.0002$ & $> 0.0006$ & $3.46^{+51.87}_{-1.56}$ & $0.22^{+0.08}_{-0.06}$  \Tstrut\Bstrut \\
		\hline
		\hline
		\multirow{5}{*}{2014}&  EMA & -\tablefootmark{*} & -  & - & - & - &  $2.62^{+0.84}_{-1.26}$ &  $0.014 \pm 0.001$ \Tstrut\Bstrut \\
		& EMB & $2.06^{+0.78}_{-0.58} \times 10^{10}$ & $11.81^{+1.82}_{-2.25}$&  $13.00^{+2.00}_{-2.35}$ &  $0.67^{+0.17}_{-0.25}$  & $2.36^{+0.93}_{-2.20}$ & $1.54^{+0.11}_{-0.10}$ & $17.41^{+0.62}_{-0.64}$ \Tstrut\Bstrut \\
		& EMC & $8.37^{+8.30}_{-5.63} \times 10^{9}$ & $4.45^{+0.02}_{-1.02}$ & $4.87^{+0.04}_{-0.95}$ & $0.94^{+0.07}_{-0.41}$  & $2.34^{+0.09}_{-1.54}$ &  $2.62^{+0.24}_{-0.21}$  &  $1.76^{+0.14}_{-0.09}$\Tstrut\Bstrut \\
		& EMD & $< 1.44 \times 10^{14}$ & $> 0.010$ & $> 0.015$ &  $> 0.0001$ & $> 0.0003$ & $0.49^{+0.47}_{-0.32}$ &  $0.11 \pm 0.01$\Tstrut\Bstrut \\
		\hline
		\multicolumn{2}{c |}{Section} & \multicolumn{3}{c |}{\ref{Subsec:Ion_Dist}}  & \multicolumn{2}{c |}{\ref{Subsec:Size_Dist}}  & \ref{Subsec:Vel_Dist}& \ref{Subsec:Vol_Dist} \Tstrut\Bstrut \\
		\hline
	\end{tabular}
\tablefoot{
\tablefoottext{*}{Unable to accurately measure density, so the distances that use $n_e$ cannot be calculated.}
\tablefoottext{$\dagger$}{$f_v = 0.1$}
} 
	\end{sidewaystable}
	
	\subsection{Thermal stability of the plasma}
	\label{Sec:Thermal_Stability}
	
\begin{figure}
		\centering
		\begin{subfigure}{1\linewidth}
			\includegraphics[width=1\linewidth]{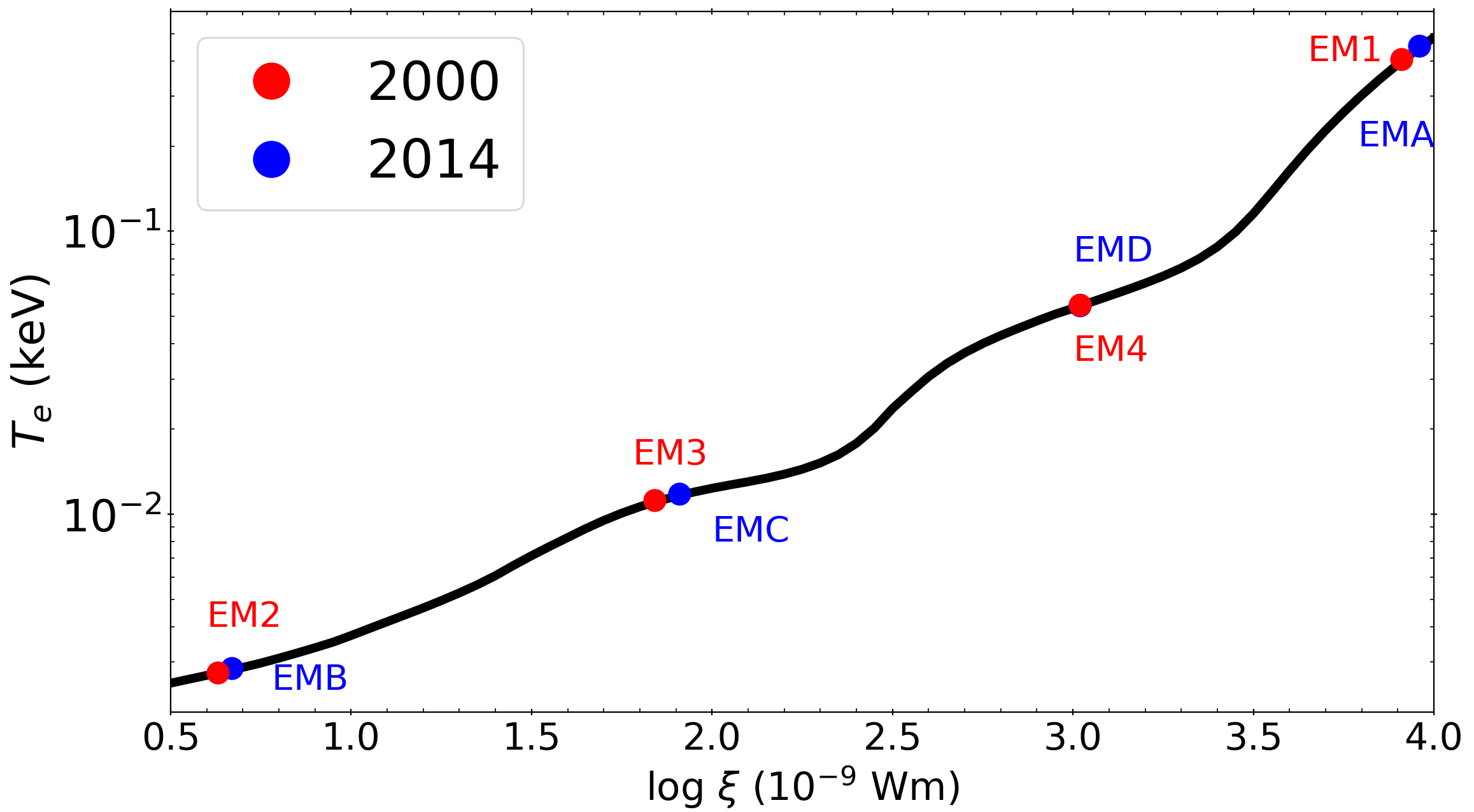}
		\end{subfigure}
		\begin{subfigure}{1\linewidth}
			\includegraphics[width=1\linewidth]{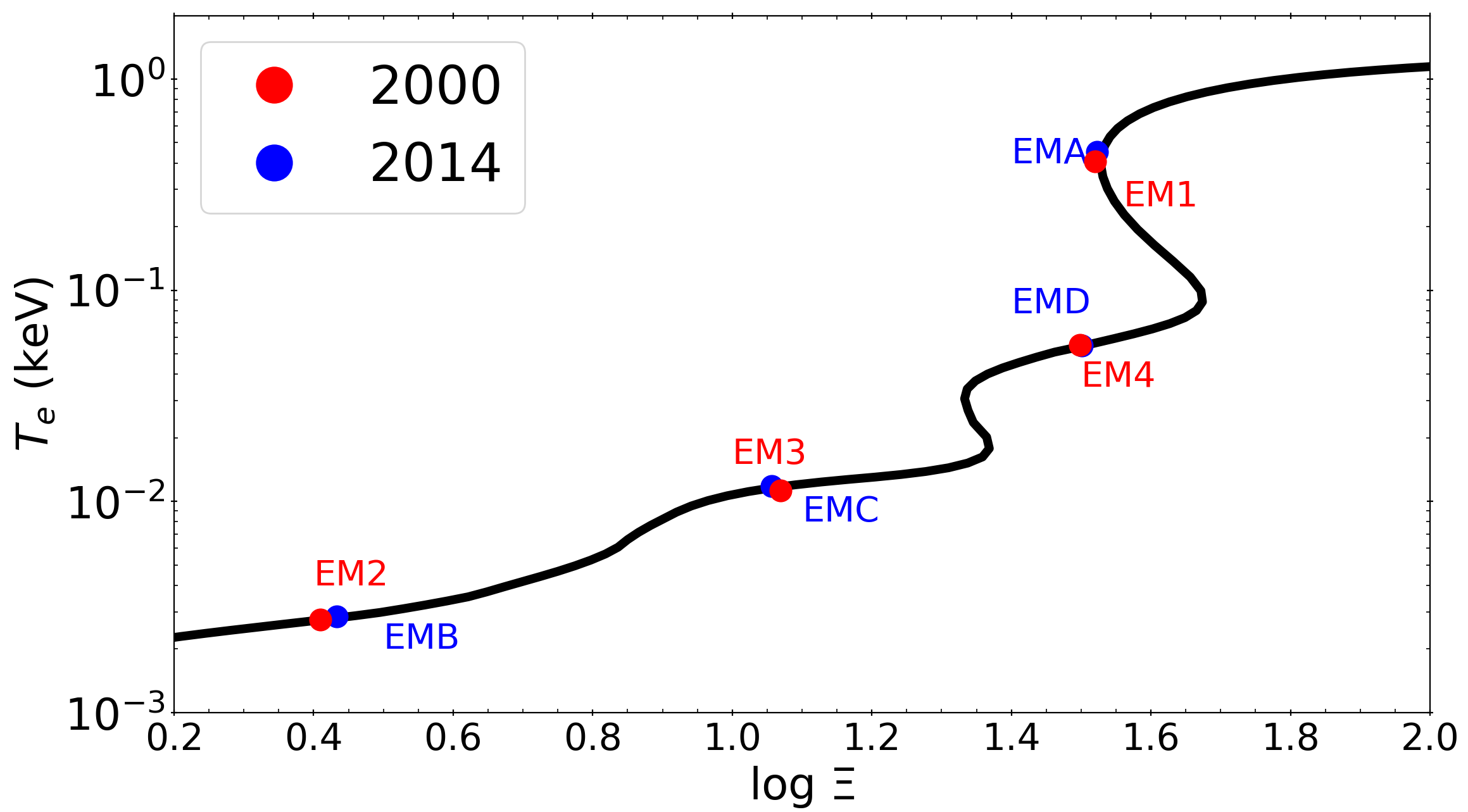}
		\end{subfigure}
		\caption{\textit{Top panel:} Electron temperature ($T_e$) of the plasma as a function of ionisation parameter ($\xi$). The labelled circles correspond to the four \texttt{PION} components for each epoch (red for 2000 and blue for 2014) at their respective ionisation parameter values from Table \ref{Table:PION_Results}. \textit{Bottom panel:} Thermal stability curve of electron temperature ($T_e$) as a function of the pressure form of the ionisation parameter ($\Xi$). On the curve, the rate of cooling equals the rate of heating. Left of the curve, cooling dominates over heating and to the right, heating dominates over cooling. The labelled circles correspond to the $\Xi$ values of the four \texttt{PION} components for each epoch: red for 2000 and blue for 2014.}
		\label{Fig:Thermal_Plots}	
	\end{figure}

	From \texttt{PION}, we are able to extract the temperature of the plasma components using ionisation and thermal balance equations, by iterating over a range of ionisation parameters, assuming photoionisation and thermal equilibrium, for each ionisation state \citep{Mehdipour2016}. Figure \ref{Fig:Thermal_Plots} displays the thermal stability of the emitting plasma, shown as a plot of electron temperature ($T_e$) as a function of either ionisation parameter ($\xi$; top panel) or the pressure form of the ionisation parameter ($\Xi$; bottom panel), within NGC 1068. The labelled circles show the locations of the four \texttt{PION} components for 2000 (red) and 2014 (blue), respectively. In the top panel of Fig. \ref{Fig:Thermal_Plots}, $T_e$ increases with an increase in $\xi$ (same for both epochs), which is to be expected as the higher the plasma ionisation, the more energy the electrons within it have, and therefore the hotter the gas is. 
	
	The bottom panel of Fig. \ref{Fig:Thermal_Plots} provides an effective tool to determine which plasma regions are thermally stable or unstable, depending on the gas being in thermal equilibrium. This plot, known as a thermal stability curve, shows the electron temperature ($T_e$) as a function of the pressure form of the ionisation parameter \citep[$\Xi$;][]{Krolik1981} which is defined as
	\begin{equation}
	\Xi = \frac{F}{n_H c k T_e} = \frac{\xi}{4\pi c k T_e},
	\label{Eq:Therm_Curve}
	\end{equation}
	where $F = L_{ion} / 4 \pi r^2$ is the flux of the ionising source (between 1 - 1000 Ryd), $n_H$ is the hydrogen number density, $c$ is the speed of light, $k$ is the Boltzmann constant, $T_e$ is the electron temperature, $\xi = \frac{L_{ion}}{n_H r^2}$ is the ionisation parameter, and $r$ is the distance of the plasma from the central SMBH. On the thermal stability curve (black line in Fig. \ref{Fig:Thermal_Plots}) the rate of heating equals the rate of cooling; to the left, cooling dominates over heating, and to the right heating dictates over cooling. Over-plotted onto the curve are the temperatures of the four \texttt{PION} components from each epoch (red for 2000 and blue for 2014), at their respective $\Xi$ values (from Eq. \ref{Eq:Therm_Curve}). EM2, EM3 and EM4 (in 2000) and EMB, EMC and EMD (in 2014) lie on the curve where the gradient is positive. This means that these components are thermally stable and can reach thermal equilibrium: should a small change in temperature occur, then the gas will move towards a heating or cooling process to balance out this change. EM1 (2000) and EMA (2014), however, lie on the curve where there is a negative gradient, although at the point where the curve starts to turn towards a positive gradient, which could be the result of an inappropriate SED we have used in the modelling. This means that the two components are thermally unstable and a little perturbation is enough to move the gas away from equilibrium. Instead, the gas may exist as two plasma phases, one cold with high density, the other hot with a low density, which together are in pressure equilibrium \citep{Krolik1981}.
	
	Components EM1 and EMA and EM4 and EMD have similar $\Xi$ values, suggesting they may be part of the same region within the outflowing wind, and are far from the locations of the lower ionised components, implying they are not associated with these. This is shown in Fig. \ref{Fig:Comp_Distances}, where the distances of EM1, EM4, EMA, and EMD are similar to each other. However, three of the distance methods for EM1 and EMA are not available, whereas three of the distance estimates for EM4 and EMD are just lower limits.
		
	
	The other two components in each epoch are possibly discrete phases, rather than possessing a continuous distribution of $\xi$, as they lie on different stable parts of the curve \citep{Ebrero2011}. Furthermore, wind instabilities will cause clumping, but the lifespan of a clump is dependent on pressure equilibrium or stability of the plasma, as shown by its location on the curve \citep[Fig. \ref{Fig:Thermal_Plots};][]{Steenbrugge2005}. Hence, EM1 and EMA cannot be in equilibrium with EM2 and EMB as their values of $\xi$ (and therefore $\Xi$) are very different. This reiterates our findings from the distances for these two components, where we can conclude that EM1 and EMA, and EM2 and EMB are not part of the same outflowing region and the ionised clouds have different properties, thus placing them at the different locations in the wind. 
	
	\cite{Bianchi2019} proposed an alternative scenario for obscured AGN where radiation pressure compression (RPC) in the outflowing gas leads to a well defined ionisation distribution and differential emission measure (DEM). However, a constant gas pressure multi-phase scenario, as adopted here, was not excluded by their results. Their alternative scenario comes from the evidence that gas surrounding the AGN is photoionised by the continuum, producing soft X-ray and optical emission from the same plasma region in AGN, suggesting high and low density gas can coexist together \citep{Bianchi2006}. 
		The AGN continuum ionises and heats up the plasma, compressing the gas and causing a gradient gas pressure. 
		Therefore, a DEM distribution is required to explain and fit the slope of a RPC gas, meaning we see X-ray emission from high-$\xi$ and low-density gas, and optical emission from low-$\xi$ and high-density gas, further away from the ionising continuum, in the same cloud \citep{Bianchi2019}.  As \cite{Bianchi2019} point out, density diagnostics with forthcoming high-throughput and high spectral resolution instrumentation will be able to distinguish between the two interpretations.

\begin{figure}
		\centering
			\includegraphics[width=1\linewidth]{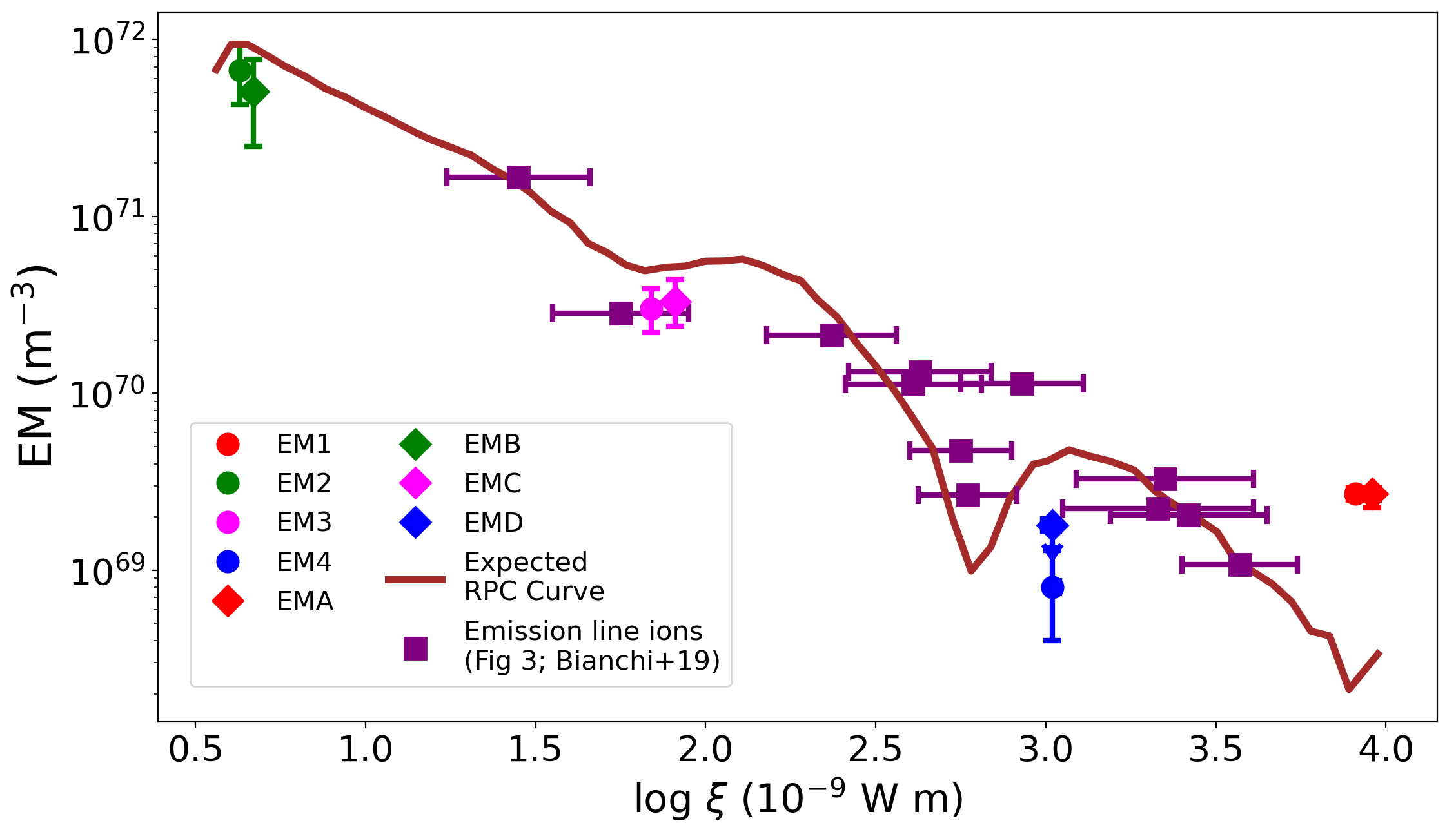}
			\caption{Emission measure (EM) of the four \texttt{PION} components in each epoch plotted versus their respective ionisation parameter ($\log \xi$). The purple squares are the ions from Fig. 3 in \cite{Bianchi2019}, whereby the EM values decrease for increasing $\log \xi$, consistent with the \texttt{PION} components (same colours and data point styles from Fig. \ref{Fig:Comp_Distances}). We also plot the expected radiative pressure compression (RPC) curve (brown line) that follows the trend of both the ions (purple squares) and \texttt{PION} components suggesting that the photoionised plasma within NGC 1068 is consistent with RPC gas.}
			\label{Fig:EM_RPC}
	\end{figure}

	In Fig. \ref{Fig:EM_RPC}, we plot the emission measure (EM) of each PION component against $\log \xi$. We then plot the 12 ions (purple squares) from Fig. 3 in \cite{Bianchi2019}, adapted for the EM axis by multiplying the DEM values by the width of each log $\xi$ bin. Finally, the DEM distribution (brown line) predicted for RPC gas is overlaid on top to test whether PIE plasma within a RPC gas is a solution. The overall trend shows a decreasing EM for an increase in $\log \xi$, consistent with the relation for a DEM ($\frac{d(EM)}{d (\log \xi)}$) in RPC \citep{Bianchi2019}. Both the distribution and overall trend suggest that the plasma is in RPC, which represents a natural solution. Here, however, the highest $\xi$ \texttt{PION} component does not fit the RPC trend, because it comes from the region that is almost transparent (almost fully ionised), so adding more high $\xi$ gas makes very little difference for the RPC solution. This can also be seen in Fig. 11 in \cite{Kallman2014}, where a plot of mass distribution as a function of $\xi$ shows an inverse relation (increase in $\xi$ for a decrease in mass). This suggests that a high $\xi$ gas has less mass and a low optical depth, and is therefore located closer to the black hole, compared to a high-density, low-$\xi$ gas, which we see in the RPC scenario.
		

	
\begin{figure}
		\centering
		\begin{subfigure}{1\linewidth}
			\includegraphics[width=1\linewidth]{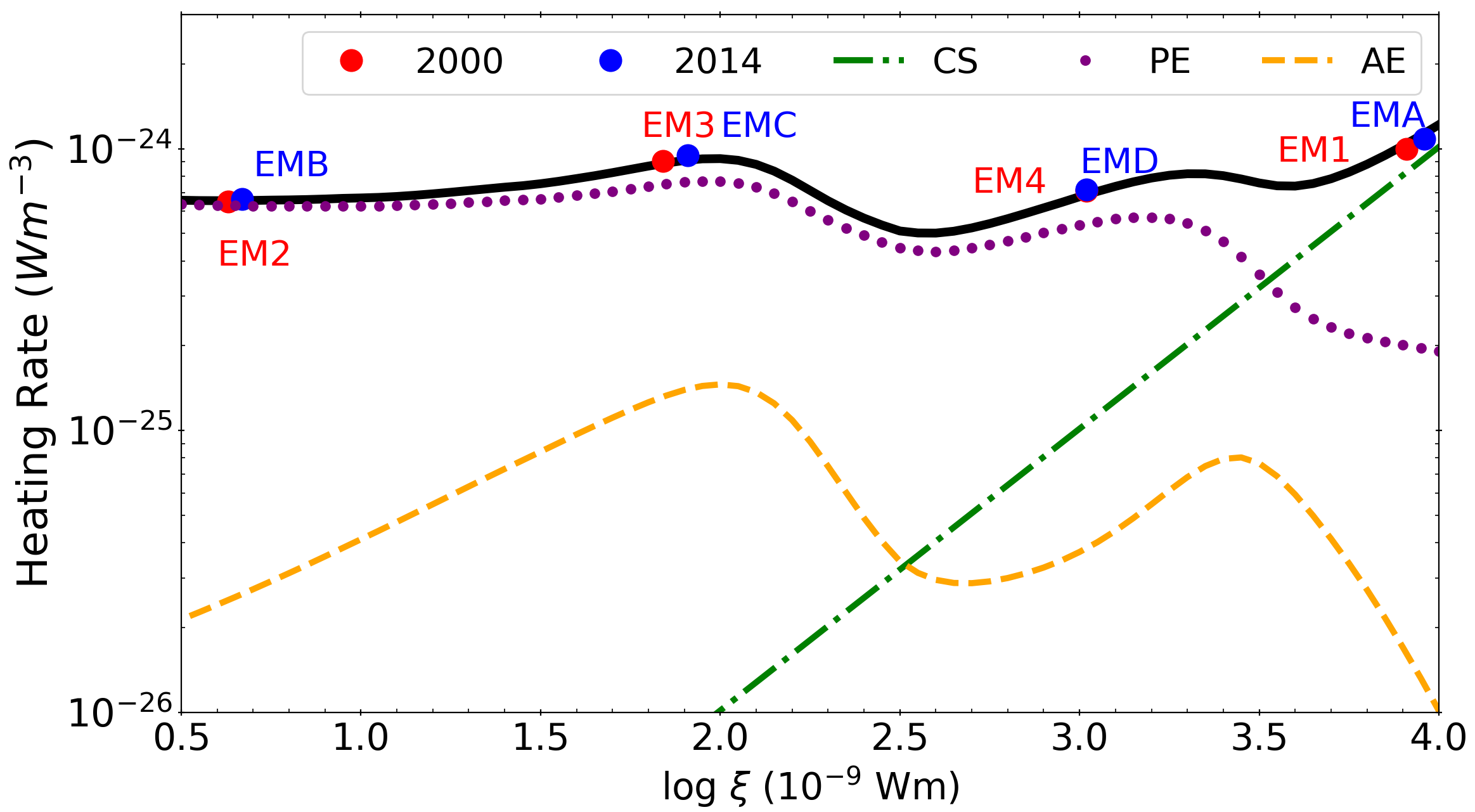}
		\end{subfigure}
		\begin{subfigure}{1\linewidth}
			\includegraphics[width=1\linewidth]{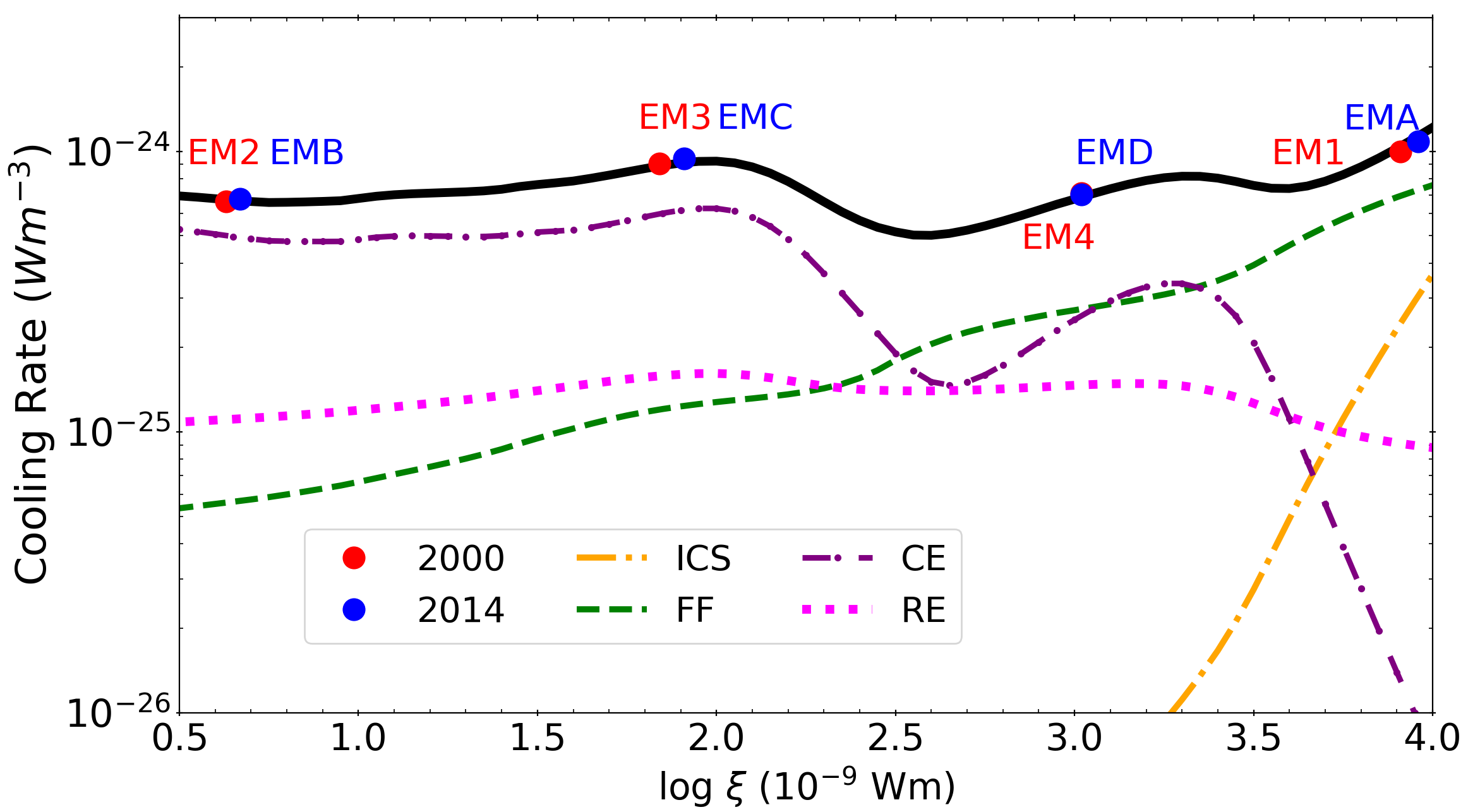}
		\end{subfigure}
		\caption{\textit{Top panel:} Total heating rate as a function of ionisation parameter ($\xi$) for NGC 1068 (black solid line). The contributions from the different heating processes are also shown: Compton scattering (CS; green dot-dashed line), photoelectrons (PE; purple dotted line), and Auger electrons (AE; orange dashed line). \textit{Bottom panel:} Total cooling rate as a function of $\xi$ for NGC 1068 (black solid line). The contributions from the individual cooling processes are also displayed: inverse-Compton scattering (ICS; orange dot-dashed line), Bremsstrahlung (FF; green dashed line), collisional excitation (CE; purple dot-dashed-dotted line), and recombination (RE; magenta dotted line). \textit{Both panels:} The labelled circles correspond to the four \texttt{PION} components in each epoch, shown as red for 2000 and blue for 2014, at their respective ionisation parameters from Table \ref{Table:PION_Results}.}
		\label{Fig:Heat_Cool_Rates}	
	\end{figure}

	In a similar way to calculating the temperatures of each component, \texttt{PION} can also calculate the rates of the various heating and cooling mechanisms, for a given ionisation parameter. Figure \ref{Fig:Heat_Cool_Rates} displays the total heating rate (top panel) and total cooling rate (bottom panel) within the PIE plasmas in NGC 1068, with the 2000 and 2014 \texttt{PION} components overplotted as red and blue circles, respectively. We also display the individual processes that make up these heating and cooling rates. The main heating processes are: Compton scattering (CS), Auger electrons (AE) and photoelectrons (PE). In Fig. \ref{Fig:Heat_Cool_Rates}, the largest contribution to heating the plasma comes from photoelectrons up to about $\log \xi = 3.5$, where Compton scattering starts dominating. This is due to X-ray ionisation increasing the number of photoelectrons in the plasma with $\log \xi < 3.5$, while above this value the free electrons interact and gain energy from the X-rays. The main cooling processes are: inverse-Compton scattering (ICS), bremsstrahlung (FF; free-free), collisional excitation (CE) and recombination (RR). The dominant cooling processes are collisional excitation for $\log \xi < 2.5$ and bremsstrahlung for $\log \xi > 2.5$. The latter comes about because the ions in the highly ionised plasma have lost all their outer shell electrons, such that when a free electron passes by, the large charge difference causes the electron to be deflected greatly, losing its kinetic energy to radiation and therefore cooling the gas. These results are consistent with the heating and cooling processes for various SEDs \citep{Mehdipour2016}.

	
	\subsection{Origin of collisionally ionised plasma}
 Finally, if CIE plasma is the origin of the 15 and 17 \AA\ lines (not considering photoexcitation), then we need to know the location of this emission. The most likely origin is from the star burst region, situated up to 1.3 kpc \citep{Garcia-Burillo2017} from the central black hole, which appears to be made up of two arcs, that together form a misshapen ring  \cite[Fig. 1a in][]{Garcia-Burillo2014}. The SBR (and CND) appears to be a dynamic structure, as shown in Fig. 10 of \cite{Garcia-Burillo2014}, where the west of the SBR is blueshifted (down to $- 180$ km s\textsuperscript{-1}), while the east is redshifted (up to $+ 180$ km s\textsuperscript{-1}). The RGS instrument has a spatial resolution ($\sim 5'$) large enough to cover the full SBR which is $0.8'$ ($\sim 50''$) in diameter, meaning we can observe the full SBR and therefore emission from the blue and redshifted sides. 
		
	Chandra LETGS images of NGC 1068 show bright X-ray emission at the nucleus (including the secondary region), but only diffuse emission further out \citep{Brinkman2002}. On the other hand, Chandra HETGS images show X-ray emission that corresponds to the star burst ring of the host galaxy, south-east of the nucleus, in the energy range between 0.8 - 1.3 keV \citep{Ogle2003}. In addition, ACIS images show X-ray emission between 0.25 - 2 keV as far out as the SBR, in both north-west and south-east directions \citep{Young2001}. The X-ray energy ranges in which emission from the SBR is observed are consistent with the RGS band where we model the \texttt{CIE} plasma, suggesting that the SBR is a good candidate for the origin of collisionally ionised emission, although direct detection is yet to be confirmed \citep[e.g.][]{Ogle2003}.

	 To confirm the legitimacy of this \texttt{CIE} component, more so than just by statistical significance in the model, we compare the calculated CIE luminosity ($L_{CIE}$) with the X-ray luminosity inferred from the far infrared (FIR) luminosity from the star burst region in NGC 1068. The star formation rate (SFR) in NGC 1068 was estimated between $0.4 - 0.7 M_{\odot}$ yr\textsuperscript{-1} \citep{Garcia-Burillo2014}. From here, we can estimate the FIR luminosity of the SFR using $L_{FIR} = \frac{SFR}{4.5 \times 10^{-44}}$ erg s\textsuperscript{-1} \citep{Kennicutt1998}: we find the FIR luminosity to be between $L_{FIR}[0.4] = 8.9 \times 10^{35}$ and $L_{FIR}[0.7] = 1.6 \times 10^{36}$ W. We then use the relation $\log \left[\frac{L_{X}}{L_{IR}}\right] > -4.3$ \citep{Symeonidis2014}, where $L_{IR}$ is the infrared luminosity and $L_X$ is the soft X-ray (0.5 - 2 keV) luminosity (as \texttt{CIE} emission is only seen in the RGS band), to infer an X-ray luminosity based on the IR luminosity. We therefore estimate a lower limit for the X-ray luminosity to be within the range $L_X > 4.46 \times 10^{31} - 7.82 \times 10^{31}$ W. The measured (0.5 - 2 keV) luminosities of the CIE components in 2000 and 2014 are $L^{2000}_{CIE} = 4.89^{+0.14}_{-0.11} \times 10^{33}$ W and $L^{2014}_{CIE} =  4.04^{+0.11}_{-0.07} \times 10^{33}$ W, respectively, which are significantly larger than the X-ray luminosities inferred from the FIR. To double check this, we use the relation $L_X = L_{FIR} 10^{-3.70}$  \citep[Eq. 8 from][]{Ranalli2003} and obtain soft X-ray luminosities of $L_X  = 1.78 \times 10^{32} - 3.19 \times 10^{32}$ W, but these are still an order of magnitude less than $L_{CIE}$.

With the CIE luminosity in both epochs being much larger than the soft X-ray luminosity limit inferred from the FIR luminosity, we conclude that either something else is contributing to the CIE X-ray luminosity, or some other process is producing this emission. The SFR values estimated from $L_{CIE}$ in both epochs, using the relation $SFR = 2.2 \times 10^{-33} L_{CIE}\ M_{\odot} yr^{-1}$ \citep[modified Eq. 14 from][]{Ranalli2003}, are $SFR_{2000} = 11\ M_{\odot} yr^{-1}$ and $SFR_{2014} = 9\ M_{\odot} yr^{-1}$. These values are significantly larger compared to the SFR from the FIR \citep{Garcia-Burillo2014}, again suggesting that the CIE plasma is unlikely to come from the SBR.
	
Alternatively, the CIE emission could originate from the  secondary region, north-east of the central black hole \citep{Brinkman2002}, assuming CIE plasma is the cause of the \ion{Fe}{XVII} lines. The secondary region is bright in both radio and X-rays, possibly due to old jet material \citep{Wilson1987}. Due to the field of view and spatial resolution of RGS, we cannot resolve the SBR and the primary and secondary regions with XMM-Newton, hence we see all the emission from each of these three regions combined. Therefore, the ionisation and outflow velocities we obtain from our \texttt{PION} modelling may also apply to emission from the secondary region, harbouring similar plasma to the primary region; however, we do not expect the location of this plasma to be as far out as the secondary region (from our distance estimates in Sect. \ref{Sec:Comp_Dist}). It is possible, therefore, that CIE plasma may be located in the secondary region if there are shocks generated in the gas at times when the nuclear jet is particularly strong.

	\section{Conclusions}
	\label{Sec:Conclusions}
	
	In this paper, we investigate the type 2 AGN NGC 1068, comparing XMM-Newton observations in 2000 and 2014, modelling the RGS and EPIC-PN spectra simultaneously. We find that the majority of the emission lines are explained by photoionised plasma, while at least two spectral features require a collisionally ionised component. The main conclusions from our analysis are detailed below. 
	
	\begin{enumerate}
		\item Four photoionised components (\texttt{PION} in \texttt{SPEX}) are required to fit the majority of the emission lines in both epochs, across the full X-ray band. We find that there is very little change in the ionisation states ($\log \xi = 1 - 4$) of these components between 2000 and 2014, implying the four plasma regions are the same. 
		
		\item We compare distance methods to get an insight into the possible locations of the emitting plasma regions within NGC 1068. Although we cannot constrain the estimates, these methods are useful in obtaining rough locations for the outflowing wind, even if there are no spectral changes between observations. 
		
		\item We investigate the thermal properties of emitting plasma regions, studying their thermal stability and heating and cooling processes. Using the thermal stability curve (Fig. \ref{Fig:Thermal_Plots}) we find that the highest ionisation state plasmas in each epoch are thermally unstable, while the other three \texttt{PION} components are stable and in thermal equilibrium. We also conclude that the highest and lowest ionisation phases cannot be part of the same regions in the outflowing wind. 
		
		\item A possible scenario for obscured AGN where radiation pressure compression (RPC) in the outflowing gas leads to a well defined ionisation distribution and differential emission measure was proposed by \cite{Bianchi2019}. The multi-phased PIE plasma modelled here follows the trend of decreasing emission measure for increasing ionisation parameter, which traces the expected slope of a RPC gas. This result suggests that the plasma surrounding the central black hole has properties consistent with a RPC gas, with X-ray and optical emissions coming from the same plasma region, depending on the ionisation and density. However, forthcoming high-throughput and high spectral resolution instrumentation will be able to distinguish between a constant gas pressure multiphased plasma, which we consider here, or one within a RPC scenario, using density diagnostics \citep{Bianchi2019}.
		
		\item From our photoionisation modelling, the ionisation parameter ($\xi$) and velocity broadening ($v_{turb}$; Table \ref{Table:PION_Results}) values, the distances (Fig. \ref{Fig:Comp_Distances}), and the thermal stability properties (Fig. \ref{Fig:Thermal_Plots}) of EM1 and EMA suggest that these highest ionisation components are very close to the central SMBH, compared to the other components. This indicates that EM1 and EMA could be part of the BLR. We cannot conclude this with certainty as the BLR is obscured by the dusty torus, although we do not fully rule out this idea as it is possible that the X-ray BLR emission is scattered into our LOS.

		\item We find evidence for emission from collisionally ionised plasma in NGC 1068, as photoionised plasma (\texttt{PION}) is unable to account for the \ion{Fe}{XVII} lines at 15 and 17 \AA. However, photoexcitation would be a more natural way to produce these lines in the outflowing gas surrounding the central black hole \citep{Kinkhabwala2002}, and the lack of \texttt{PION} emission in these lines may be due to incomplete information regarding photoexcitation in \texttt{SPEX}. However, if CIE emission was the cause of the Fe-L lines, 
			we find no conclusive evidence for the CIE emission originating from the SBR or the secondary region. This suggests that CIE may not be the underlying emission component to produce these lines, but instead only statistically accounts for the lines \texttt{PION} cannot produce, where photoexcitation is not accounted for.

	\end{enumerate}
	NGC 1068 has changed very little, if at all, in the 14 year period between XMM-Newton observations. Future monitoring of NGC 1068 would be important in assessing if this ionisation state of the plasma is maintained, or if this AGN is dynamic in a similar way to type 1 AGN. Multiwavelength observations are paramount in establishing as many intrinsic properties as possible of this obscured AGN in order to constrain and develop the underlying SED for accurate photoionisation modelling.
	
	\begin{acknowledgements}
		
		We would like to thank Jelle Kaastra for his helpful conversions and the anonymous referee for their useful and insightful comments.
		
		This work is based on observations obtained with XMM-Newton, an ESA science mission with instruments and contributions directly funded by ESA Member States and NASA.
		
		S.G.W. acknowledges the support of a PhD studentship awarded by the UK Science \& Technology Facilities Council (STFC).
		
		M.M. is supported by the Netherlands Organisation for Scientific Research (NWO) through the Innovational Research Incentives Scheme Vidi grant 639.042.525. 
		
		S.B. acknowledges financial support from the Italian Space Agency under grant ASI-INAF 2017-14-H.O.

	\end{acknowledgements}
	

\begin{thebibliography}{}
\bibitem[Antonucci \& Miller(1985)]{Antonucci1985} Antonucci, R. R. J. \& Miller, J. S. 1985, ApJ, 297, 621
\bibitem[Bauer et al.(2015)]{Bauer2015} Bauer, F. E., Arévalo, P., Walton, D. J., et al. 2015, ApJ, 812, 116
\bibitem[Behar et al.(2017)]{Behar2017} Behar, E., Peretz, U., Kriss, G. A., et al. 2017, A\&A, 601, A17
\bibitem[Behar et al.(2003)]{Behar2003} Behar, E., Rasmussen, A. P., Blustin, A. J., et al. 2003, ApJ, 598, 232
\bibitem[Beuchert et al.(2017)]{Beuchert2017} Beuchert, T., Markowitz, A. G., Dauser, T., et al. 2017, A\&A, 603, A50
\bibitem[Bianchi et al.(2006)]{Bianchi2006} Bianchi, S., Guainazzi, M., \& Chiaberge, M. 2006, A\&A, 448, 499
\bibitem[Bianchi et al.(2019)]{Bianchi2019} Bianchi, S., Guainazzi, M., Laor, A., Stern, J., \& Behar, E. 2019, MNRAS, 485, 416
\bibitem[Bianchi et al.(2001)]{Bianchi2001} Bianchi, S., Matt, G., \& Iwasawa, K. 2001, MNRAS, 322, 669
\bibitem[Blustin et al.(2007)]{Blustin2007} Blustin, A. J., Kriss, G. A., Holczer, T., et al. 2007, A\&A, 466, 107
\bibitem[Blustin et al.(2005)]{Blustin2005} Blustin, A. J., Page, M. J., Fuerst, S. V., Brand uardi-Raymont, G., \& Ashton, C. E. 2005, A\&A, 431, 111
\bibitem[Brinkman et al.(2002)]{Brinkman2002} Brinkman, A. C., Kaastra, J. S., van der Meer, R. L. J., et al. 2002, A\&A, 396, 761
\bibitem[Cappi et al.(2016)]{Cappi2016} Cappi, M., De Marco, B., Ponti, G., et al. 2016, A\&A, 592, A27
\bibitem[Cash(1979)]{Cash1979} Cash, W. 1979, ApJ, 228, 939
\bibitem[Crenshaw et al.(2010)]{Crenshaw2010} Crenshaw, D. M., Schmitt, H. R., Kraemer, S. B., Mushotzky, R. F., \& Dunn, J. P. 2010, ApJ, 708, 419
\bibitem[den Herder et al.(2001)]{denHerder2001} den Herder, J. W., Brinkman, A. C., Kahn, S. M., et al. 2001, A\&A, 365, L7
\bibitem[Di Gesu et al.(2017)]{DiGesu2017} Di Gesu, L., Costantini, E., Piconcelli, E., et al. 2017, A\&A, 608, A115
\bibitem[Ebrero et al.(2011)]{Ebrero2011} Ebrero, J., Kriss, G. A., Kaastra, J. S., et al. 2011, A\&A, 534, A40
\bibitem[García-Burillo et al.(2016)]{Garcia-Burillo2016} García-Burillo, S., Combes, F., Ramos Almeida, C., et al. 2016, ApJ, 823, L12
\bibitem[García-Burillo et al.(2014)]{Garcia-Burillo2014} García-Burillo, S., Combes, F., Usero, A., et al. 2014, A\&A, 567, A125
\bibitem[García-Burillo et al.(2017)]{Garcia-Burillo2017} García-Burillo, S., Viti, S., Combes, F., et al. 2017, A\&A, 608, A56
\bibitem[Grafton-Waters et al.(2020)]{Grafton-Waters2020} Grafton-Waters, S., Branduardi-Raymont, G., Mehdipour, M., et al. 2020, A\&A, 633, A62
\bibitem[Gu et al.(2019)]{Gu2019} Gu, L., Raassen, A. J. J., Mao, J., et al. 2019, A\&A, 627, A51
\bibitem[Guainazzi et al.(1999)]{Guainazzi1999} Guainazzi, M., Matt, G., Antonelli, L. A., et al. 1999, MNRAS, 310, 10
\bibitem[Huchra et al.(1999)]{Huchra1999} Huchra, J. P., Vogeley, M. S., \& Geller, M. J. 1999, The Astrophysical Journal Supplement Series, 121, 287
\bibitem[Imanishi et al.(2018)]{Imanishi2018} Imanishi, M., Nakanishi, K., Izumi, T., \& Wada, K. 2018, ApJ, 853, L25
\bibitem[Jethwa et al.(2015)]{Jethwa2015} Jethwa, P., Saxton, R., Guainazzi, M., Rodriguez-Pascual, P., \& Stuhlinger, M. 2015, A\&A, 581, A104
\bibitem[Kaastra(2017)]{Kaastra2017} Kaastra, J. S. 2017, A\&A, 605, A51
\bibitem[Kaastra et al.(2016)]{KB2016} Kaastra, J. S. \& Bleeker, J. A. M. 2016, A\&A, 587, A151
\bibitem[Kaastra et al.(2012)]{Kaastra2012} Kaastra, J. S., Detmers, R. G., Mehdipour, M., et al. 2012, A\&A, 539, A117
\bibitem[Kaastra et al.(2014)]{Kaastra2014} Kaastra, J. S., Kriss, G. A., Cappi, M., et al. 2014, Science, 345, 64
\bibitem[Kaastra et al.(1996)]{Kaastra1996} Kaastra, J. S., Mewe, R., \& Nieuwenhuijzen, H. 1996, in UV and X-ray Spectroscopy of Astrophysical and Laboratory Plasmas, 411–414
\bibitem[Kaastra et al.(2002)]{Kaastra2002} Kaastra, J. S., Steenbrugge, K. C., Raassen, A. J. J., et al. 2002, A\&A, 386, 427
\bibitem[Kalberla et al.(2005)]{Kalberla2005} Kalberla, P. M. W., Dedes, L., Arnal, E. M., et al. 2005, 331, 81
\bibitem[Kallman et al.(2014)]{Kallman2014} Kallman, T., Evans, D. A., Marshall, H., et al. 2014, ApJ, 780, 121
\bibitem[Kennicutt (1998)]{Kennicutt1998} Kennicutt, Robert C., J. 1998, ARA\&A, 36, 189
\bibitem[Kinkhabwala et al.(2002)]{Kinkhabwala2002} Kinkhabwala, A., Sako, M., Behar, E., et al. 2002, ApJ, 575, 732
\bibitem[Kollatschny \& Zetzl(2013)]{Kollatschny2013} Kollatschny, W. \& Zetzl, M. 2013, A\&A, 558, A26
\bibitem[Kraemer et al.(2015)]{Kraemer2015} Kraemer, S. B., Sharma, N., Turner, T. J., George, I. M., \& Crenshaw, D. M. 2015, ApJ, 798, 53
\bibitem[Krolik et al.(1981)]{Krolik1981} Krolik, J. H., McKee, C. F., \& Tarter, C. B. 1981, ApJ, 249, 422
\bibitem[Liedahl et al.(1990)]{Liedahl1990} Liedahl, D. A., Kahn, S. M., Osterheld, A. L., \& Goldstein, W. H. 1990, ApJ, 350, L37
\bibitem[Lodders et al.(2009)]{Lodders2009} Lodders, K., Palme, H., \& Gail, H. P. 2009, Landolt Börnstein, 4B, 712
\bibitem[Magdziarz \& Zdziarski(1995)]{Magdziarz1995} Magdziarz, P. \& Zdziarski, A. A. 1995, MNRAS, 273, 837
\bibitem[Mao et al.(2018)]{Mao5548} Mao, J., Kaastra, J. S., Mehdipour, M., et al. 2018, A\&A, 612, A18
\bibitem[Mao et al.(2019)]{Mao3783} Mao, J., Mehdipour, M., Kaastra, J. S., et al. 2019, A\&A, 621, A99
\bibitem[Marinucci et al.(2016)]{Marinucci2016} Marinucci, A., Bianchi, S., Matt, G., et al. 2016, MNRAS, 456, L94
\bibitem[Matt (2002)]{Matt2002} Matt, G. 2002, MNRAS, 337, 147
\bibitem[Matt et al.(2004)]{Matt2004} Matt, G., Bianchi, S., Guainazzi, M., \& Molendi, S. 2004, A\&A, 414, 155
\bibitem[Mehdipour et al.(2018)]{Mehdipour2018} Mehdipour, M., Kaastra, J. S., Costantini, E., et al. 2018, A\&A, 615, A72
\bibitem[Mehdipour et al.(2016)]{Mehdipour2016} Mehdipour, M., Kaastra, J. S., \& Kallman, T. 2016, A\&A, 596, A65
\bibitem[Mehdipour et al.(2017)]{Mehdipour2017} Mehdipour, M., Kaastra, J. S., Kriss, G. A., et al. 2017, A\&A, 607, A28
\bibitem[Middei et al.(2018)]{Middei2018} Middei, R., Bianchi, S., Cappi, M., et al. 2018, A\&A, 615, A163
\bibitem[Miller \& Antonucci(1983)]{Miller1983} Miller, J. S. \& Antonucci, R. R. J. 1983, ApJ, 271, L7
\bibitem[M\"{u}ller S\'{a}nchez et al.(2009)]{Muller-Sanchez2009} M\"{u}ller S\'{a}nchez, F., Davies, R. I., Genzel, R., et al. 2009, ApJ, 691, 749
\bibitem[Netzer(2013)]{Netzer2013} Netzer, H. 2013, The Physics and Evolution of Active Galactic Nuclei
\bibitem[Ogle et al.(2003)]{Ogle2003} Ogle, P. M., Brookings, T., Canizares, C. R., Lee, J. C., \& Marshall, H. L. 2003, A\&A, 402, 849
\bibitem[Osterbrock(1991)]{Osterbrock1991} Osterbrock, D. E. 1991, Reports on Progress in Physics, 54, 579
\bibitem[Panessa et al.(2006)]{Panessa2006} Panessa, F., Bassani, L., Cappi, M., et al. 2006, A\&A, 455, 173
\bibitem[Peretz et al.(2019)]{Peretz2019} Peretz, U., Miller, J. M., \& Behar, E. 2019, ApJ, 879, 102
\bibitem[Petrucci et al.(2013)]{Petrucci2013} Petrucci, P. O., Paltani, S., Malzac, J., et al. 2013, A\&A, 549, A73
\bibitem[Porquet \& Dubau(2000)]{Porquet2000} Porquet, D. \& Dubau, J. 2000, A\&AS, 143, 495
\bibitem[Pounds \& Vaughan(2006)]{Pounds2006} Pounds, K. \& Vaughan, S. 2006, MNRAS, 368, 707
\bibitem[Rahin \& Behar(2020)]{RB2020} Rahin, R. \& Behar, E. 2020, ApJ, 904, 40
\bibitem[Ramos Almeida \& Ricci(2017)]{Ricci2017} Ramos Almeida, C. \& Ricci, C. 2017, Nature Astronomy, 1, 679
\bibitem[Ranalli et al.(2003)]{Ranalli2003} Ranalli, P., Comastri, A., \& Setti, G. 2003, A\&A, 399, 39
\bibitem[Sako et al.(2000)]{Sako2000} Sako, M., Kahn, S. M., Paerels, F., \& Liedahl, D. A. 2000, ApJ, 543, L115
\bibitem[Schurch et al.(2004)]{Schurch2004} Schurch, N. J., Warwick, R. S., Griffiths, R. E., \& Kahn, S. M. 2004, MNRAS, 350, 1
\bibitem[Semena et al.(2019)]{Semena2019} Semena, A. N., Sazonov, S. Y., \& Krivonos, R. A. 2019, Astronomy Letters, 45, 490
\bibitem[Snedden \& Gaskell(1999)]{Snedden1999} Snedden, S. A. \& Gaskell, C. M. 1999, ApJ, 521, L91
\bibitem[Steenbrugge et al.(2005)]{Steenbrugge2005} Steenbrugge, K. C., Kaastra, J. S., Crenshaw, D. M., et al. 2005, A\&A, 434, 569
\bibitem[Str\"{u}der et al.(2001)]{Struder2001} Str\"{u}der, L., Briel, U., Dennerl, K., et al. 2001, A\&A, 365, L18
\bibitem[Symeonidis et al.(2014)]{Symeonidis2014} Symeonidis, M., Georgakakis, A., Page, M. J., et al. 2014, MNRAS, 443, 3728
\bibitem[Titarchuk(1994)]{Titarchuk1994} Titarchuk, L. 1994, ApJ, 434, 570
\bibitem[Viti et al.(2014)]{Viti2014} Viti, S., García-Burillo, S., Fuente, A., et al. 2014, A\&A, 570, A28
\bibitem[Wakker(2006)]{Wakker2006} Wakker, B. P. 2006, The Astrophysical Journal Supplement Series, 163, 282
\bibitem[Whewell et al.(2015)]{Whewell2015} Whewell, M., Branduardi-Raymont, G., Kaastra, J. S., et al. 2015, A\&A, 581, A79
\bibitem[Wilson \& Ulvestad(1987)]{Wilson1987} Wilson, A. S. \& Ulvestad, J. S. 1987, ApJ, 319, 105
\bibitem[Young et al.(2001)]{Young2001} Young, A. J., Wilson, A. S., \& Shopbell, P. L. 2001, ApJ, 556, 6	
	\end{thebibliography}

\end{document}